\documentclass[useAMS,usenatbib]{mn2e}
\usepackage{graphicx}
\usepackage{txfonts}

\title[Redshift evolution of oxygen abundances]
      {The metallicity -- redshift relations for emission-line SDSS galaxies: 
       examination of the dependence on the star formation rate}

\author[L.S.~Pilyugin et al.]
       {L.S.~Pilyugin$^{1,2}$,
        M.A.~Lara-L\'{o}pez$^{3}$,
        E.K.~Grebel$^{2}$,
        C.~Kehrig$^4$,
        I.A.~Zinchenko$^1$,  
\newauthor    \'{A}.R.~L\'{o}pez-S\'{a}nchez$^{3,5}$,
        J.M.~V\'{\i}lchez$^{4}$, 
        L.~Mattsson$^{6}$  \\ 
       $^{1}$ Main Astronomical Observatory
             of National Academy of Sciences of Ukraine,
             27 Zabolotnogo str., 03680 Kiev, Ukraine \\
       $^{2}$ Astronomisches Rechen-Institut, Zentrum f\"{u}r Astronomie 
             der Universit\"{a}t Heidelberg, 
             M\"{o}nchhofstr.\ 12--14, 69120 Heidelberg, Germany \\
       $^{3}$ Australian Astronomical Observatory, PO BOX 915, North Ryde, NSW 1670, Australia\\
       $^{4}$ Instituto de Astrof\'{\i}sica de Andaluc\'{\i}a,
             CSIC, Apdo, 3004, 18080 Granada, Spain \\
       $^{5}$ Department of Physics and Astronomy, Macquarie University, NSW 2109, Australia\\       
       $^{6}$ DARK Cosmology Centre, Niels Bohr Institute,
            University of Copenhagen, Juliane Maries Vej 30,
            DK-2100, Copenhagen \O, Denmark\\
              }

\date{Accepted 2013 March 26. Received 2013 March 03; in original form 2013 January 11}

\begin{document}

\maketitle

\begin{abstract}
We analyse the oxygen abundance and specific star formation rates ($sSFR$) variations 
with redshift in star-forming SDSS galaxies of different masses. 
We find that the maximum value of the $sSFR$, $sSFR_{max}$, 
decreases when the stellar mass, $M_S$, of a galaxy increases, and decreases with decreasing of redshift. 
The $sSFR_{max}$ can exceed the time-averaged $sSFR$ by about an order of magnitude for 
massive galaxies.  
The metallicity -- redshift relations for subsamples of galaxies with $sSFR$ = $sSFR_{max}$ and 
with $sSFR$ = 0.1$\times$$sSFR_{max}$ coincide for massive (log($M_S$/$M_{\odot}$) $\ga$ 10.5, with stellar mass $M_S$ 
in solar units) galaxies and differ for low-mass galaxies. 
This suggests that there is no correlation between oxygen abundance and $sSFR$ in massive galaxies
and that the  oxygen abundance correlates with the $sSFR$ in low-mass galaxies.  
We find evidence in favour of that the irregular galaxies 
show, on average, higher $sSFR$ and lower oxygen abundances than the spiral 
galaxies of similar masses and that the mass -- metallicity relation 
for spiral galaxies differs slightly from that for irregular galaxies. 
The fact that our sample of low-mass galaxies is the mixture of spiral and irregular
galaxies can be responsible for the dependence of the  metallicity -- redshift relation 
on the $sSFR$ observed for the low-mass SDSS galaxies.
The mass -- metallicity and luminosity -- metallicity relations obtained for irregular 
SDSS galaxies agree with corresponding relations for nearby irregular galaxies with direct abundance determinations. 
We find that the aperture effect does not make a significant contribution to the redshift variation of oxygen 
abundances in SDSS galaxies.
\end{abstract}


\section{Introduction}

It is well established that the oxygen abundance correlates with 
galaxy mass (or luminosity), O/H = $f$($M_S$), in the sense that the higher the 
galaxy mass (luminosity), the higher the heavy element content 
\citep[][among others]{Lequeuxetal1979,Garnettshields1987, Skillmanetal1989,Vilacostas1992,
Zaritskyetal1994,Pilyugin2001,Gusevaetal2009,LopezSanchezEsteban2010AA517}.  
In recent years, the number of available spectra of emission-line galaxies 
has increased dramatically due to the completion of several 
large spectral surveys, in particular the Sloan Digital Sky Survey (SDSS) \citep{Yorketal2000} . 
Measurements of emission lines in SDSS spectra have been carried out 
for abundance determinations and investigation of the luminosity -- 
metallicity relation \citep[e.g.][]{Tremontietal2004}.   
It is discussed whether the mass -- metallicity relation
can be caused by the galaxy-downsizing effect, where star formation ceases
in high-mass galaxies at earlier times and shifts to lower-mass
galaxies at later epochs \citep{Cowieetal1996}.
The study of oxygen abundance evolution with redshift in  
emission-line SDSS galaxies provides 
strong evidence in favour of the galaxy-downsizing effect
\citep{Thuanetal2010,LaraLopezetal2010AA519,PilyuginThuan2011}.

It is important to note there are different absolute
O/H scales. The empirical metallicity scale is defined by the direct method 
($T_e$ method) and the empirical calibrations based on it 
\citep[e.g.][]{Pilyugin2001AA369,PettiniPagel2004MNRAS348,PilyuginThuan2005ApJ631,Pilyuginetal2010ApJ720}. 
Metallicities derived using calibrations based on photoionisation models 
\citep[e.g.][]{McGaugh1991ApJ380,KewleyDopita2002ApJS142,KobulnickyKewley2004ApJ617} 
tend to be, systematically higher ($\sim$0.4 dex or even more) than those derived using the direct method 
\citep[see reviews by][]{LopezSanchezEsteban2010AA517,LopezSanchezetal2012MNRAS426}. 
Therefore, there are large discrepancies between oxygen abundances in the SDSS galaxies 
obtained in different works using different calibrations.

It has been argued that the mass -- metallicity 
relation is due to a more general relation in the 3-dimensional space formed by the stellar mass, 
gas-phase oxygen abundance and the star formation rate. 
This 3-dimensional structure is referred to as a Fundamental Plane (FP) by 
\citet{LaraLopezetal2010AA521}, suggesting that the stellar mass can be expressed as 
a function of the $SFR$ and O/H, i.e. $M_S$ = $f(SFR,$O/H). On the other hand,  
\citet{Mannuccietal2010} describes this 3-dimensional space as a surface, 
the Fundamental Metallicity Relation (FMR) in which the O/H = $f(M_s,SFR)$. For a further 
discussion of the 3-dimensional structure of this space see  
\citet{Yatesetal2012,LaraLopezetal2012arXiv1207.0950L}.

It is also known that the mass -- metallicity relation changes with redshift $z$.
Thus, one can expect that in general the oxygen abundance in a galaxy 
correlates with three parameters, i.e., O/H = $f$($M_S$, $SFR$, $z$). 
Study of the redshift evolution of oxygen abundances in 
galaxies of different masses may tell us something about the evolution 
of galaxies and provide important constrains on the models for the chemical 
evolution of galaxies. 
In this paper, we will examine whether the metallicity -- redshift relations for 
SDSS galaxies of different masses  are dependent on the star formation rate.

The paper is structured as follows.
Selection criteria for the sample of the SDSS galaxies with reliable abundance 
determinations are reported in Section 2. In Section 3 we establish the redshift evolution
of the maximum star formation rates in galaxies of different masses as well  
the redshift evolutions of oxygen abundances in galaxies of different masses and $sSFR$. 
The properties of irregular and low-mass spiral galaxies are compared  in Section 4.
In section 5 we discuss whether  the aperture effect makes 
a significant contribution to the redshift variation of oxygen 
abundances in the SDSS galaxies. We also estimate the time-averaged values of the $sSFR$
in massive galaxies. 
The conclusions are given in Section 6. 

Throughout the paper, we will be using the following notations for the line fluxes,
\begin{equation}
R_2 =  I_{\rm \rm [OII] \lambda 3727+ \lambda 3729} /I_{\rm {\rm H}\beta },
\end{equation}
\begin{equation}
N_2  = I_{\rm \rm [NII] \lambda 6548+ \lambda 6584} /I_{\rm {\rm H}\beta },
\label{equation:n2standard}
\end{equation}
\begin{equation}
S_2  = I_{\rm \rm [SII] \lambda 6717 + \lambda 6731} /I_{\rm {\rm H}\beta },
\end{equation}
\begin{equation}
R_3  = I_{\rm {\rm [OIII]} \lambda 4959 + \lambda 5007} /I_{\rm {\rm H}\beta }.
\label{equation:r3standard}
\end{equation}
The stellar masses of galaxies $M_S$ are expressed in the solar mass units.

\section{Data base}

\subsection{A sample of SDSS galaxies}

\begin{figure*}
\resizebox{1.00\hsize}{!}{\includegraphics[angle=000]{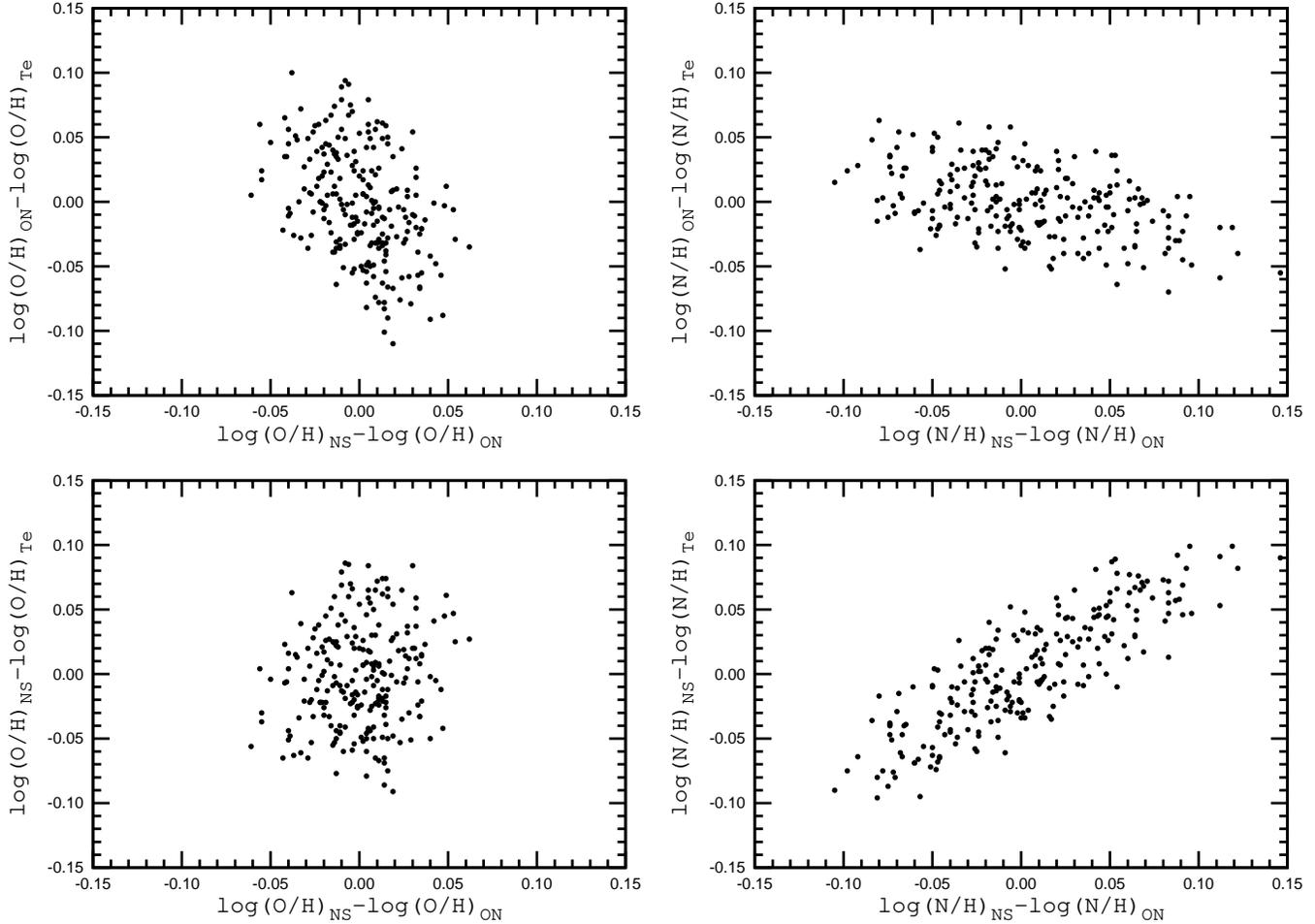}}
\caption{
Comparison of differences between oxygen (nitrogen) abundances 
determined in different ways for the sample of reference  H\,{\sc ii} 
regions ($E$ sample). 
}
\label{figure:e-e}
\end{figure*}

Our study is based on SDSS data. 
Line flux measurements in SDSS spectra have been carried out by several 
groups. We use here the data made publicly available  
by the MPA/JHU group\footnote{The catalogs are available in the SDSS 
DR9 database (tables galSpecExtra, galSpecInfo and galSpecLine) 
at http://www.sdss3.org/dr9/}.  These data catalogs give   
line flux measurements, redshifts and various other derived physical properties  
such as stellar masses for a large sample of SDSS galaxies.
The techniques used to construct the catalogues are described 
in \citet{Brinchmannetal2004,Tremontietal2004}
and various other publications by the same authors. 

In a first step, we extract all emission-line objects with measured fluxes in the 
H$\beta$, H$\alpha$, [O\,{\sc ii}]$\lambda \lambda$3727,3729, 
[O\,{\sc iii}]$\lambda$5007, [N\,{\sc ii}]$\lambda$6584, [S\,{\sc ii}]$\lambda$6717 and 
[S\,{\sc ii}]$\lambda$6731 emission lines.  
The hydrogen, oxygen, nitrogen and sulfur lines 
will serve to estimate oxygen and nitrogen abundances relative to hydrogen. 
The wavelength range of the SDSS spectra is 3800 -- 9300 \AA\ so that
for nearby galaxies with redshift z $\la$ 0.023, the 
[O\,{\sc ii}]$\lambda$3727+$\lambda$3729 emission line is out of that range. 
Therefore all SDSS galaxies with  z $\la$ 0.023 were excluded. 
In the other end, distant galaxies with redshift z $\ga$ 0.33, the 
[S\,{\sc ii}]$\lambda$6717+$\lambda$6731 emission line is out of that range. 
In general, reliable estimates of the oxygen abundance in a SDSS galaxy can be 
found even if the [O\,{\sc ii}]$\lambda$3727+$\lambda$3729 emission line 
is not available \citep{PilyuginThuan2007ApJ669,PilyuginMattsson2011MNRAS412} or if the 
[S\,{\sc ii}]$\lambda$6717+$\lambda$6731 emission line is not measured 
\citep{Pilyuginetal2010ApJ720}. However,  our strategy to construct a sample 
of the SDSS galaxies with rather reliable oxygen abundances 
requires both these lines. 
Thus, all galaxies in our total sample have redshifts greater than   
$\sim$ 0.023 and lower than $\sim$ 0.33.
The redshift $z$ and stellar mass M$_S$ of each galaxy were also taken from 
the MPA/JHU catalogs.

The emission-line fluxes are then corrected for interstellar 
reddening using the theoretical H$\alpha$/H$\beta$ ratio and the 
\citet{Whitford1958}  interstellar reddening law 
due to \citet{Izotovetal1994}. We adopt case $B$ recombination with a 
density of 100 cm$^{-3}$, and an electronic temperature of 10$^4$ K, 
which yields a theoretical ratio  of H$\alpha$/H$\beta$=2.86 \citep{Osterbrock2006}. 
In several cases, the derived 
value of the extinction $C$(H$\beta$) is negative and has then been set to zero.

The [O\,{\sc iii}]$\lambda$4959 line is required to obtain the $R_3$ value. 
However, this line is measured with much less accuracy than [O\,{\sc iii}]$\lambda$5007 line.
Since the [O\,{\sc iii}]$\lambda$5007 and $\lambda$4959 lines originate from transitions from the 
same energy level, their flux ratio is due only to the transition probability ratio, 
which is  very close to 3 \citep{Storey2000}. Therefore, the value of $R_3$ can be estimated as
\begin{equation}
R_3  = 1.33 \times \mbox{\rm [O\,{\sc iii}]$\lambda$5007/H$\beta$}
\label{equation:o3}   
\end{equation}
Similarly,  the value of $N_2$ is estimated without the line 
[N\,{\sc ii}]$\lambda$6548. The [N\,{\sc ii}]$\lambda$6584 and $\lambda$6548 lines 
also originate from transitions from the 
same energy level and the transition probability ratio for those lines is again
close to 3 \citep{Storey2000}. The value of $N_2$ can therefore be estimated as 
\begin{equation}
N_2  = 1.33 \times \mbox{\rm [N\,{\sc ii}]$\lambda$6584/H$\beta$}
\label{equation:n2}   
\end{equation}
The spectroscopic data so assembled form the basis of the present study. 

The intensities of strong, easily measured lines can be used to separate  
different types of emission-line objects according to their main  
excitation mechanism. \citet{Baldwinetal1981}  proposed a diagram 
(the so called BPT classification diagram) where 
the excitation properties of H\,{\sc ii} regions are characterised by 
plotting the low-excitation [N\,{\sc ii}]$\lambda$6584/H$\alpha$ 
line ratio against the high-excitation [O\,{\sc iii}]$\lambda$5007/H$\beta$ 
line ratio. The exact location of the dividing line between 
H\,{\sc ii} regions and AGNs is still controversial, however 
\citep[see, e.g.,][]{Kewleyetal2001,Kauffmannetal2003,Stasinskaetal2006}. 
Here we adopt the dividing line suggested by \citet{Kauffmannetal2003}
\begin{equation}
\log (\mbox{\rm [O\,{\sc iii}]$\lambda$5007/H$\beta$}) =
\frac{0.61}{\log (\mbox{\rm [N\,{\sc ii}]$\lambda$6584/H$\alpha$})-0.05} +1.3,
\label{equation:kauff}   
\end{equation}
which separates objects with H\,{\sc ii} spectra from those containing an AGN.
By this formula, the suspected AGNs have been excluded from further consideration.

\subsection{Abundances}

A new method (labelled "the $C$ method") for oxygen and nitrogen 
abundance determination has recently been suggested \citep{Pilyuginetal2012MNRAS424}. 
It is based on the standard assumption
that H\,{\sc ii} regions with similar strong-line intensities have similar physical properties and abundances. 
From a sample of reference H\,{\sc ii} regions with well-measured ($T_{\rm e}$-based) abundances 
we choose a counterpart for the considered H\,{\sc ii} region by comparison of four combinations of 
strong-line intensities:  $P$ = $R_3$/($R_2$ + $R_3$) 
(excitation parameter), log$R_3$, log($N_2$/$R_2$), and log($S_2$/$R_2$)
The oxygen and nitrogen abundances, as well as the electron temperature in the 
studied H\,{\sc ii} region may then be assumed to be the same as in its counterpart.  
To get more reliable abundances, we select a number of reference H\,{\sc ii} regions with abundances near those
in the counterpart H\,{\sc ii} region and then derive the abundance in the studied H\,{\sc ii} region through extra-/interpolation. 

It has been suggested that reliable oxygen and nitrogen abundances in 
H\,{\sc ii} region can be derived even if the [S\,{\sc ii}]$\lambda$6717+$\lambda$6731 emission line 
is not measured \citep{Pilyuginetal2010ApJ720} or if the [O\,{\sc ii}]$\lambda$3727+$\lambda$3729 
emission line is not available \citep{PilyuginMattsson2011MNRAS412}.
Hence, the $C$ method can be modified in the following way. 
To find the counterpart for the H\,{\sc ii} region under study, we will compare not 
the four values that are expressed in terms of the strong line intensities 
but two sets of three values: 1) log$R_3$,  $P$ and  log($N_2$/$R_2$) and 
2) log$R_3$, log$N_2$ and log($N_2$/$S_2$). The oxygen and nitrogen abundances 
determined with the help of the counterparts selected with the former set of values will 
be referred to as (O/H)$_{C_{\rm ON}}$ and (N/H)$_{C_{\rm ON}}$, while the oxygen and nitrogen abundances 
determined with the help of the counterparts selected with the latter set of values will 
be referred to as (O/H)$_{C_{\rm NS}}$ and (N/H)$_{C_{\rm NS}}$.  
Using the collected data with $T_e$-based abundances from  
\citet{Pilyuginetal2012MNRAS424} we select a sample of reference H\,{\sc ii} regions 
for which all the absolute differences for oxygen abundances 
(O/H)$_{C_{\rm ON}}$  -- (O/H)$_{T_{e}}$ and (O/H)$_{C_{\rm NS}}$  -- (O/H)$_{T_{e}}$
and for nitrogen abundances 
(N/H)$_{C_{\rm ON}}$  -- (N/H)$_{T_{e}}$ and (N/H)$_{C_{\rm NS}}$  -- (N/H)$_{T_{e}}$ are less 0.1 dex.
This sample of reference H\,{\sc ii} regions contains 233  objects and 
will be used for abundance determinations in the following. This sample will be referred to as $E$ sample (etalon sample) below. 
Hence, here we use the empirical metallicity scale defined by the H\,{\sc ii} regions 
with abundances derived through the direct method ($T_e$ method).   
Comparison between the oxygen abundances (O/H)$_{C_{NS}}$ and (O/H)$_{C_{ON}}$ and
nitrogen abundances (N/H)$_{C_{NS}}$ and (N/H)$_{C_{ON}}$) provides a selection criterion
for star-forming galaxies with accurate line fluxes measurements (see below). 

The SDSS spectra of distant galaxies are closer to global spectra of galaxies 
than spectra of single  H\,{\sc ii} regions. \citet{Pilyuginetal2012MNRAS421} have Monte Carlo 
simulated global spectra of composite nebulae as a mix of spectra of individual components, 
based on spectra of well-studied H\,{\sc ii} regions in nearby galaxies. Abundance 
analysis of the artificial composite nebulae yielded the conclusion that oxygen 
and nitrogen abundances determined using the ON and NS calibrations are in good 
agreement with each other and near (within $\sim$0.2 dex) the mean H$\beta$ luminosity-weighted 
value of oxygen and nitrogen abundances of individual components of the composite nebula.

\subsection{Selection criteria}

\begin{figure*}
\resizebox{1.00\hsize}{!}{\includegraphics[angle=000]{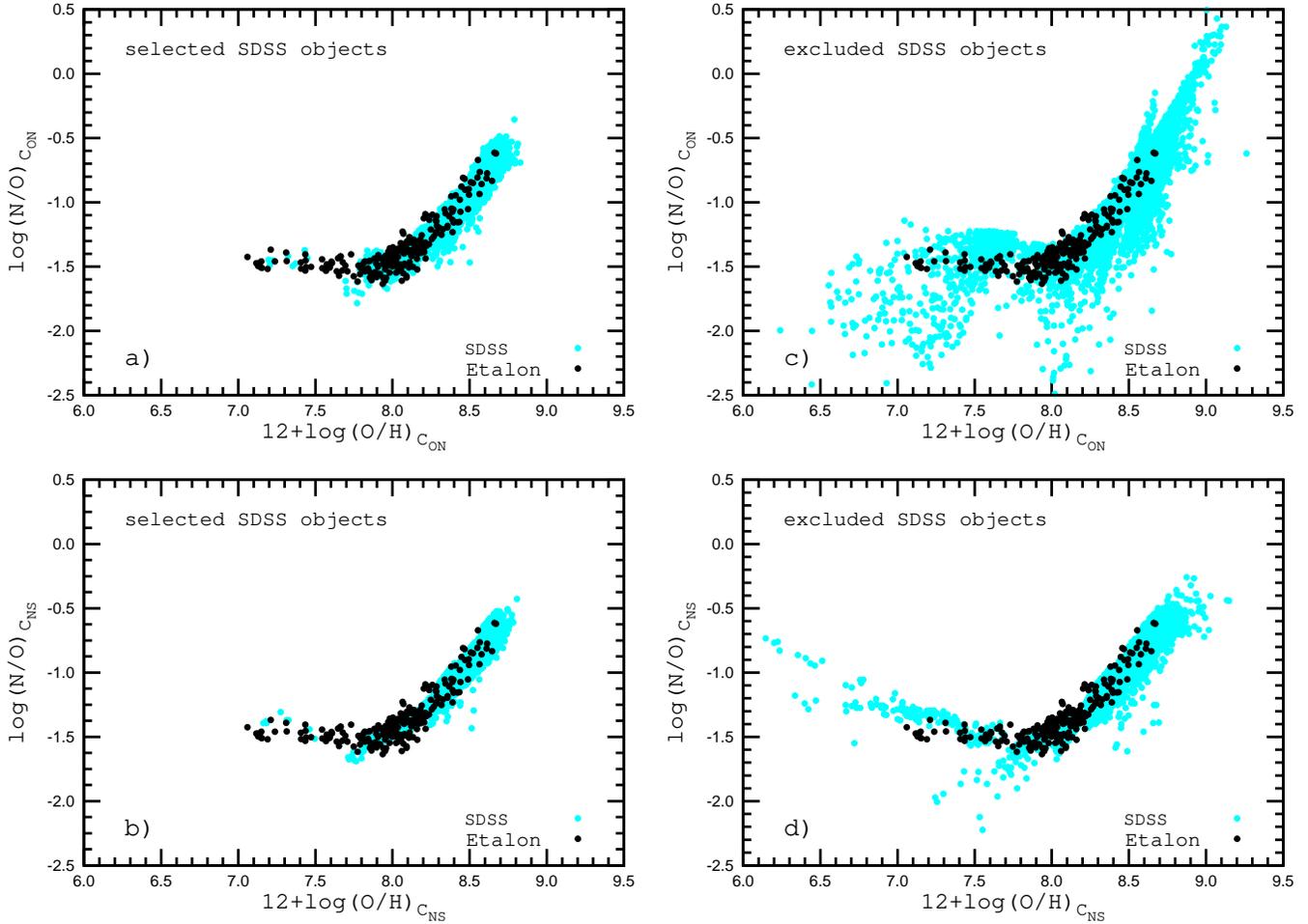}}
\caption{ 
The N/O -- O/H diagram.  
The gray (blue) points in panel $a)$ show the (X/H)$_{C_{ON}}$ abundances 
for the selected sample of the SDSS galaxies, 
the gray (blue) points in  panel $b)$ show the (X/H)$_{C_{NS}}$ abundances. 
The gray (blue) points in panels $c)$ and $d)$ show the same for the rejected SDSS objects. 
The dark (black) points in the all panels show the $T_e$-based abundances 
for the sample of reference  H\,{\sc ii} regions ($E$ sample). 
(A colour version of this figure is available in the online version.) 
}
\label{figure:oh-nosdss}
\end{figure*}

Our study is based on SDSS spectra. 
To study the redshift evolution of oxygen and nitrogen abundances, 
accurate oxygen and nitrogen abundance determinations are of course a prerequisite.
Abundances in H\,{\sc ii} regions determined 
from SDSS spectra suffer from the following problem, however. 
Line fluxes in SDSS spectra are 
measured by an automatic procedure, which inevitably introduces large flux
errors for some objects. We therefore wish to exclude them from further consideration. 
 A  method to recognise such objects relies on the fact   
that there exist a fundamental H\,{\sc ii} region sequence. 
The emission line properties of photoionised H\,{\sc ii} regions are governed 
by their heavy element content and by the electron temperature distribution 
within each photoionised nebula. 
The latter is controlled by the ionising star cluster's spectral 
energy distribution and by the chemical composition of the H\,{\sc ii} region. 
The evolution of a giant extragalactic H\,{\sc ii} region associated with a  
star cluster is thus caused by a gradual change over time (due to stellar evolution) 
of the integrated stellar energy distribution. This phenomenon has been studied in numerous 
investigations
\citep[][among others]{Stasinska1978AA66,Stasinska1980AA84,McCalletal1985ApJS57,DopitaEvans1986ApJ307,
Moyetal2001AA365,StasinskaIzotov2003AA397,Dopitaetal2006ApJS167}. 
The general conclusion from those studies is that H\,{\sc ii} regions 
ionised by star clusters form a well-defined fundamental sequence in 
different emission-line diagnostic diagrams. 
The existence of such a fundamental sequence forms the basis for
studying various properties of extragalactic H\,{\sc ii} regions.
First, \citet{Baldwinetal1981} have suggested the position of an 
object in some well-chosen 
emission-line diagrams can be used to separate H\,{\sc ii} regions 
ionised by star clusters from other types of emission-line objects. 
Second, \citet{Pageletal1979MNRAS189} and \citet{Alloinetal1979AA78} have suggested that the positions 
of H\,{\sc ii} regions in some emission-line diagrams can be calibrated in 
terms of their oxygen abundances. 

Following \citet{Thuanetal2010} and \citet{Mannuccietal2010} we adopt here a simple method to recognise 
star-forming galaxies with accurate line fluxes measurements.
It is based on the idea  
that if {\it i)} an object belongs to the fundamental H\,{\sc ii} region sequence,
and {\it ii)} its line fluxes are measured accurately, then the different 
relations between the line fluxes and the physical characteristic of 
H\,{\sc ii} regions, 
based on different emission lines, should yield similar  physical characteristics (such as electron 
temperatures and abundances) of that object. 

Fig.~\ref{figure:e-e} show comparisons of differences between values of oxygen 
(nitrogen) abundances determined in various ways for the $E$ sample of reference 
H\,{\sc ii} regions. Fig.~\ref{figure:e-e} also shows that the 
correlation between differences in oxygen abundances determined in different ways is weak.
A correlation between differences in nitrogen abundances determined in different ways
is more evident. Although these correlations are  
not very tight one can expect that the conditions  
$|$log(O/H)$_{C_{NS}}$ -- log(O/H)$_{C_{ON}}$$|$ $\le$ 0.05 and 
$|$log(N/H)$_{C_{NS}}$ -- log(N/H)$_{C_{ON}}$$|$ $\le$ 0.10 allow to select the 
SDSS objects with more or less reliable abundances.  
Thus, to select star-forming galaxies with accurate line fluxes measurements, 
we require that, for each galaxy, the oxygen (O/H)$_{C_{NS}}$ and (O/H)$_{C_{ON}}$ 
and nitrogen (N/H)$_{C_{NS}}$ and (N/H)$_{C_{ON}}$ abundances agree.

It should be noted that the $C$ method was developed for   H\,{\sc ii} regions 
in the low-density regime. 
Therefore we include only objects with reliable electron densities $n_e$.  
The electron density is estimated from the sulfur line ratio 
[S\,{\sc ii}]$\lambda$6717/[S\,{\sc ii}]$\lambda$6731.
According to the  five-level atom solution for the S$^{+}$ ion, 
the [S\,{\sc ii}]$\lambda$6717/[S\,{\sc ii}]$\lambda$6731 is a reliable 
indicator of the electron density in the interval from $n_e$ $\sim$ 10$^2$ cm$^{-3}$ 
to  $n_e$ $\sim$ 10$^4$ cm$^{-3}$
\begin{equation}
\log\,n_e  =  8.448 - 15.101 x + 14.419 x^2 - 5.115 x^3
\label{equation:ne}   
\end{equation}
where $x$ = [S\,{\sc ii}]$\lambda$6717/[S\,{\sc ii}]$\lambda$6731. 
We exclude objects with 
[S\,{\sc ii}]$\lambda$6717/[S\,{\sc ii}]$\lambda$6731 $<$ 1.2, i.e., those with 
$n_e$ $\ga$ 170 cm$^{-3}$. 
At densities $n_e$ $\la$ 10$^2$ cm$^{-3}$  the ratio 
[S\,{\sc ii}]$\lambda$6717/[S\,{\sc ii}]$\lambda$6731 is almost independent 
on the electron density and reaches the limit value 1.43.
The measured ratio [S\,{\sc ii}]$\lambda$6717/[S\,{\sc ii}]$\lambda$6731 in some 
objects is in excess of this limit value, that can be considered as evidence in 
favour of that the line measurements in the SDSS spectra of this object  
are not accurate. Therefore the objects with 
[S\,{\sc ii}]$\lambda$6717/[S\,{\sc ii}]$\lambda$6731 $>$ 1.55 (expected error in 
the line measurements is larger $\sim$10\%) were also excluded. 

Another subset of galaxies were excluded from consideration using the following conditions. 
In some cases, H\,{\sc ii} regions in giant spiral galaxies seem to 
be catalogued as dwarf galaxies of small dimensions. 
Thus, if a "galaxy" has a Petrosian radius less than 3 kpc, this galaxy is then excluded from our list. 
Further, the aperture correction factor (see below) for some galaxies is  $<1$,
i.e., the galaxy light within the fiber exceeds the total light of the galaxy. 
The galaxy light fraction within the fiber is less than 0.7 for the bulk of 
galaxies. Therefore, the objects with the galaxy light fraction within the fiber 
higher 0.8 were also excluded from further consideration. 

Our final sample contains 98.986 SDSS objects. The O/H -- N/O diagram 
can tell us something about the credibility of our selection criteria and 
about the reliability of the abundance determinations.
Fig.~\ref{figure:oh-nosdss} displays the N/O -- O/H diagram.  
The gray (blue points) in the panel $a)$ show the (X/H)$_{C_{ON}}$ abundances 
for the selected sample of the SDSS galaxies, 
the gray (blue points) in the panel $b)$ show the (X/H)$_{C_{NS}}$ abundances.
The dark (black) points (in both panels) show the $T_e$-based abundances 
for the sample of reference  H\,{\sc ii} regions ($E$ sample). 
A quick look at Fig.~\ref{figure:oh-nosdss} immediately tells us that the selected 
SDSS galaxies, with very 
few exceptions, are located within the area outlined by the 
the reference H\,{\sc ii} regions with $T_e$-based abundances (as in the 
case (X/H)$_{C_{ON}}$ and as in the case (X/H)$_{C_{NS}}$ abundances in 
the SDSS galaxies). 
This suggests that the strategy outlined above allows us to construct a sample 
of SDSS galaxies with rather reliable oxygen and nitrogen abundances. 
 
Panels $c)$ and $d)$ in  Fig.~\ref{figure:oh-nosdss} show the 
N/O -- O/H diagrams for the rejected SDSS objects.
The positions of rejected objects occupy a much larger area  in 
the (N/O)$_{C_{ON}}$ -- (O/H)$_{C_{ON}}$ (and (N/O)$_{C_{NS}}$ -- (O/H)$_{C_{NS}}$)  
diagram than that outlined by the reference H\,{\sc ii} regions, although some 
rejected objects are located within that area. 
The weak point of our selection criterion is that 
we lose some fraction of objects with reliable  (O/H)$_{C_{ON}}$ abundances 
if only  sulfur [S\,{\sc ii}]$\lambda$6717 and [S\,{\sc ii}]$\lambda$6731 lines measurement  
involve errors, and some fraction of objects with reliable  (O/H)$_{C_{NS}}$ 
if only  oxygen [O\,{\sc ii}]$\lambda \lambda$3727,3729 emission lines measurement 
involve errors.

The resultant reduced sample is the basis for our study. 
For clarity and simplicity,  only (O/H)$_{C_{ON}}$ abundances will be considered 
in the following.

\section{The redshift variations of SFR and oxygen abundances in SDSS galaxies}

\begin{figure*}
\resizebox{0.950\hsize}{!}{\includegraphics[angle=000]{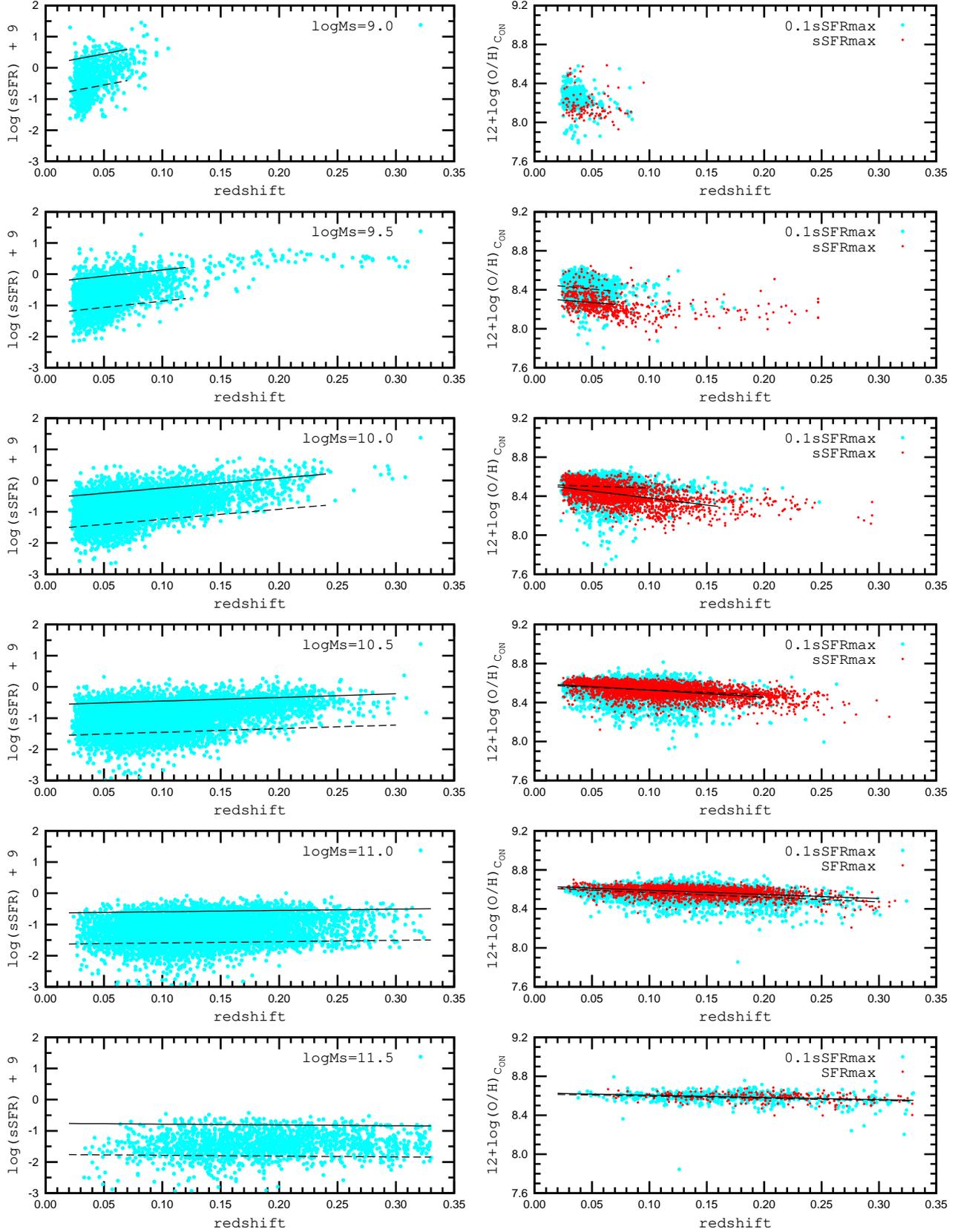}}
\caption{{\it Left  panels.}
Redshift dependence of the $sSFR$ in galaxies of different masses. 
Light-gray (blue) points are individual galaxies. The solid line is the  "$sSFR$$_{max}$ -- 
redshift" relation, while the dashed line show the "0.1$\times$$sSFR$$_{max}$ -- 
redshift" relation.
{\it Right panels.} 
Redshift dependence of the oxygen abundance in galaxies of different masses 
and different $sSFR$. 
Dark-gray (red) points are galaxies with log($sSFR$)=log($sSFR_{max}$)$\pm$0.2. The solid line is 
the best fit to those data. The light-gray (blue) pints 
are galaxies with   log($sSFR$)=log(0.1$\times$$sSFR_{max}$)$\pm$0.2, while the dashed line is 
the best fit to the data. 
(A colour version of this figure is available in the online version.) 
}
\label{figure:z-sfr}
\end{figure*}

\subsection{Specific star formation rate}

The specific star formation rate $sSFR$ is defined as the $SFR$ per unit  
stellar mass in a galaxy, i.e., $sSFR$ = $SFR$/$M_S$. 
In estimating $SFR$s from the H$\alpha$ luminosity we adopt the calibration  
suggested by \citet{Kennicutt1998ARAA36} 
\begin{equation}
SFR \;\;({\rm M_{\odot}} {\rm  yr^{-1}})  = \frac{L_{H{\alpha}} \;\;({\rm W})}{1.27 \times 10^{34} } ,
\label{equation:sfr}   
\end{equation}
where 
\begin{equation}
L_{H{\alpha}} = 4{\pi}d^2F_{H{\alpha}}^{cor},
\label{equation:lha}   
\end{equation}
is the H$\alpha$ luminosity. In equation (\ref{equation:lha}) $d$ is the distance to a galaxy, and $F_{H{\alpha}}^{cor}$ is 
the H$\alpha$ flux corrected for extinction and aperture effect.

The reddening-corrected  H$\beta$ flux $F_{\rm {\rm H}\beta}^O$ is obtained from 
the observed  $F_{\rm {\rm H}\beta}^{obs}$ flux using the relation  \citep{Blagrave2007} 
\begin{equation}
\log(F_{\rm {\rm H}\beta}^{obs}/F_{\rm {\rm H}\beta}^O) = - C_{\rm \rm H\beta} ,  
\label{equation:hbcorr}   
\end{equation}
where  $C$(H$\beta$) is  the extinction coefficient. 
The generic reddening-corrected flux ratio is assumed to be $F_{\rm {\rm H}\alpha}^O$/$F_{\rm {\rm H}\beta}^O$ = 2.86. 

In addition to the extinction correction, the H$\alpha$ luminosity also requires
an aperture correction to account for the fact that only a limited amount of
emission from a galaxy is detected through the 3" diameter fiber. We 
estimate the aperture correction factor $A$ using  
the aperture correction method from \citet{Hopkinsetal2003ApJ599}
\begin{equation}
A = 10^{-0.4(r_{Petro} - r_{fiber})} , 
\label{equation:a}   
\end{equation}
where $r_{fiber}$ and $r_{Petro}$ are $r$-band fiber and Petrosian magnitudes. 
This correction is based on the assumption that the emission line flux can be traced overall by the 
 $r$-band emission. 
The aperture-corrected  H$\alpha$ flux is $F_{\rm {\rm H}\alpha}^{cor}$ = $A$ $\times$ $F_{\rm {\rm H}\alpha}^O$. 

Distances to galaxies were calculated from 
\begin{equation}
d = \frac{cz}{H_0} ,
\label{equation:d}   
\end{equation}
where $d$ is the distance in Mpc,  $c$ the speed of light in km s$^{-1}$, and $z$ the redshift. 
$H_0$ is the Hubble constant, here assumed to be equal to  72 ($\pm$8) km s$^{-1}$ Mpc$^{-1}$ 
\citep{Freedman2001}.

\subsection{The redshift variations  of $sSFR$s and oxygen abundances in galaxies of different masses} 

Studies of the redshift evolution of the $SFR$s in the SDSS galaxies 
encounter selection-effect problems in the sense that the weak 
hydrogen emission line luminosity cannot be measured in the distant objects. 
The lower limit of the measured H$\beta$ luminosity at the redshift $z$=0.25 
is three orders of magnitude higher than that at $z$=0.01 (see, for example, Fig.7 in 
\citet{Pilyuginetal2012MNRAS419}). To avoid this selection effect we will consider 
primarily the redshift evolution of the maximum value of the specific star formation rate 
$sSFR_{max}$ in galaxies of different masses.

We consider six stellar-mass bins centered
on the values log$M_s$ = 9.0, 9.5, 10.0, 10.5, 11.0, 11.5 with interval width +/- 0.2 dex. 
The left panels in  Fig.~\ref{figure:z-sfr} show  
the redshift variations of $sSFR$s in galaxies of different masses. 
Light-gray (blue) points are individual galaxies. The solid line is the  linear 
"$sSFR_{max}$ -- redshift" relation. This relation has been obtained as follows. 
As a first step, the average value of the $sSFR$ has to be found. 
In a second step, the galaxies with $sSFR$s lower than the average value have to be 
rejected and a new average $sSFR$ value has to be obtained. 
In a third  step, the galaxies with the $sSFR$s lower than the new average value is 
rejected and a "$sSFR$ -- redshift" relation is then obtained. 
The fourth step is to reject the galaxies with the $sSFR$s below this relation 
and thus a final "$sSFR_{max}$ -- redshift" relation 
has been obtained. In the deriving the final "$sSFR_{max}$ -- redshift" relation 
the objects with deviations exceeding the mean value of deviations by a factor three 
were excluded. Hence, the final "$sSFR_{max}$ -- redshift" is determined using 
around 5-10\% of galaxies in a fixed mass interval. In the first and second steps 
we use the average  $sSFR$ value rather than  "$sSFR$ -- redshift" relation 
in order to exclude the objects with low $sSFR$s in an attempt to overcome the selection effect.
The dashed lines in the  left panels in  Fig.~\ref{figure:z-sfr} show 
the "0.1$\times$$sSFR_{max}$ -- redshift" relations.

The left panels in  Fig.~\ref{figure:z-sfr} shows 
that the maximum specific star formation rate correlates 
with the stellar mass of a galaxy: the value of the $sSFR_{max}$ 
decreases with increasing of the $M_S$. This nicely confirms  the galaxy downsizing effect, 
where star formation ceases in high-mass galaxies at earlier times and  
shifts to lower-mass galaxies at later epochs \citep{Cowieetal1996}. 

The right panels in  Fig.~\ref{figure:z-sfr} show the redshift variations  
of the oxygen abundance in galaxies of different masses and different $sSFR$s. 
Dark-gray (red) points are galaxies with $sSFR$s close to $sSFR_{max}$, i.e., galaxies which 
are located within +/-0.2 dex along the solid line in the 
corresponding left panel in  Fig.~\ref{figure:z-sfr}.
The solid line is the linear best fit to those data. The light-gray (blue) pints are galaxies with  
 $sSFR$s close to 0.1$\times$$sSFR_{max}$, i.e., galaxies which 
are located within the band of width +/-0.2 dex along the dashed line in the 
corresponding left panel in  Fig.~\ref{figure:z-sfr}.
The dashed line is the best fit to those data. 

A closer look at Fig.~\ref{figure:z-sfr} reveals that the metallicity -- redshift relations for 
massive galaxies (log$M_s$ $\ga$ 10.5) do not depend on the $sSFR$.
While the $sSFR$ changes by an order of magnitude (from $sSFR_{max}$ to  0.1$\times$$sSFR_{max}$), 
the metallicity -- redshift relations for those galaxies are similar. 
This, we interpret as evidence that there is no correlation between oxygen abundance and $sSFR$ in massive galaxies.  
The metallicity -- redshift relation for low-mass galaxies (log$M_s$ $\la$ 10.5) on the other hand, depends on 
the $sSFR$: the galaxies with highest $sSFR$ show lowest oxygen abundances. 

\subsection{Comparison to a previous analysis of SFR-dependence of the O/H -- $M_s$ relation} 
 
The mass -- metallicity relation of galaxies as a function of (specific) star formation rate
has previously been considered in several different studies \citep{Ellisonetal2008,
LaraLopezetal2010AA521,Mannuccietal2010,Yatesetal2012,AndrewsMartini2013aph,LaraLopezetal2012arXiv1207.0950L}.
It is difficult to directly compare those results with ours
for two reasons. First, our method for abundance determination differs from that 
used in other studies. Second, a single mass -- metallicity relation for galaxies from a
rather large redshift range were constructed in the papers cited above, i.e., 
the redshift evolution of galaxies is not taken into account. 
Therefore, we will limit ourselves to comparison of the general picture of the 
$SFR$-dependence of the mass -- metallicity relation.
But first we review the results obtained by previous investigators.

\citet{Ellisonetal2008} considered a mass -- metallicity relation for 
SDSS galaxies, binned in specific star formation rate.  
Metallicities were obtained using the calibration from \citet{KewleyDopita2002ApJS142}. 
They found that at high stellar masses (log$M_s$ $\ga$ 10) the 
mass -- metallicity relation exhibits no dependence on the $sSFR$. At lower stellar masses, 
there is a tendency for galaxies with higher $sSFR$ to have lower metallicities for a 
given stellar mass. The offset in metallicity from highest to lowest $sSFR$ bins is 
$\sim$ 0.10-0.15 dex in the stellar mass range 9 $\la$ log$M_s$ $\la$ 10.

\citet{Mannuccietal2010} studied the relation between stellar mass,
metallicity and star formation rate for SDSS galaxies with redshifts  
0.07 $\la$ $z$ $\la$ 0.30. The oxygen abundances were derived using
calibrations based either on the [N\,{\sc ii}]$\lambda$6584/H$\alpha$ or on 
$R_{23}$ = ([O\,{\sc ii}]$\lambda$3727 + [O\,{\sc iii}]$\lambda$4959,5007)/H$\beta$ 
parameter. When both calibrations can be used, galaxy metallicity is defined as the 
average of these two values. Moreover, they selected only galaxies where the two values 
of metallicity differ no more than 0.25 dex.  
They found that galaxies with high stellar masses (log$M_s$ $\ga$ 10.9) show no 
correlation between metallicity and star formation rate, while at low stellar masses more 
active galaxies also show lower metallicity.

\citet{Yatesetal2012}  studied the relation between stellar mass,
metallicity and star formation rate for SDSS galaxies with redshifts  
0.005 $\la$ $z$ $\la$ 0.25. Two metallicity-data sets were used.
The oxygen abundances were derived following \citet{Mannuccietal2010} in the first case 
and using the catalogued values in SDSS-DR7 in the second case. 
They found that the metallicity is dependent on 
star formation rate for a fixed mass, but that the trend is opposite for galaxies in the low and 
high stellar-mass ends. Low-mass galaxies with high $SFR$ are more metal poor than
quiescent low-mass galaxies. Massive galaxies have lower metallicity if their 
star formation rates are small.
\citet{LaraLopezetal2012arXiv1207.0950L} have reached a similar conclusion.
 
\citet{AndrewsMartini2013aph} considered SDSS galaxies with redshifts
0.027 $\la$ $z$ $\la$ 0.25. It has been assumed that galaxies at a given stellar mass, 
or simultaneously a given stellar mass and star formation rate, have the same metallicity.  
The SDSS galaxies were stacked in bins of (1) stellar mass and (2) both stellar 
mass and star formation rate. The auroral lines were measured in stacked spectra and 
the abundances were derived using the direct method.  They found that the $SFR$-dependence 
appeared monotonic with stellar mass.

In summary, the general conclusion in previous studies, is that the mass -- metallicity 
relation is dependent of (specific) star formation rate at low stellar masses
\citep{Ellisonetal2008,Mannuccietal2010,Yatesetal2012,AndrewsMartini2013aph,LaraLopezetal2012arXiv1207.0950L}. 
Our results for low-mass galaxies are in agreement with these results.
The conclusions by various authors regarding the $SFR$-dependence of the mass -- metallicity 
relation at high stellar masses are controversial, however. 
On the one hand, \citet{Ellisonetal2008} and \citet{Mannuccietal2010} have found that 
at high stellar masses the mass -- metallicity relation shows no dependence on 
(specific) star formation rate. 
Our results for massive galaxies are in agreement with the conclusions by  \citet{Ellisonetal2008} 
and  \citet{Mannuccietal2010}. On the other hand, 
\citet{Yatesetal2012,AndrewsMartini2013aph,LaraLopezetal2012arXiv1207.0950L}  
have found that at high stellar masses the mass -- metallicity relation depends on (specific) 
star formation rate. 
Our results for massive galaxies are  in conflict with those results.

\section{Low-mass galaxies: spirals vs irregulars}

\begin{figure*}
\resizebox{1.00\hsize}{!}{\includegraphics[angle=000]{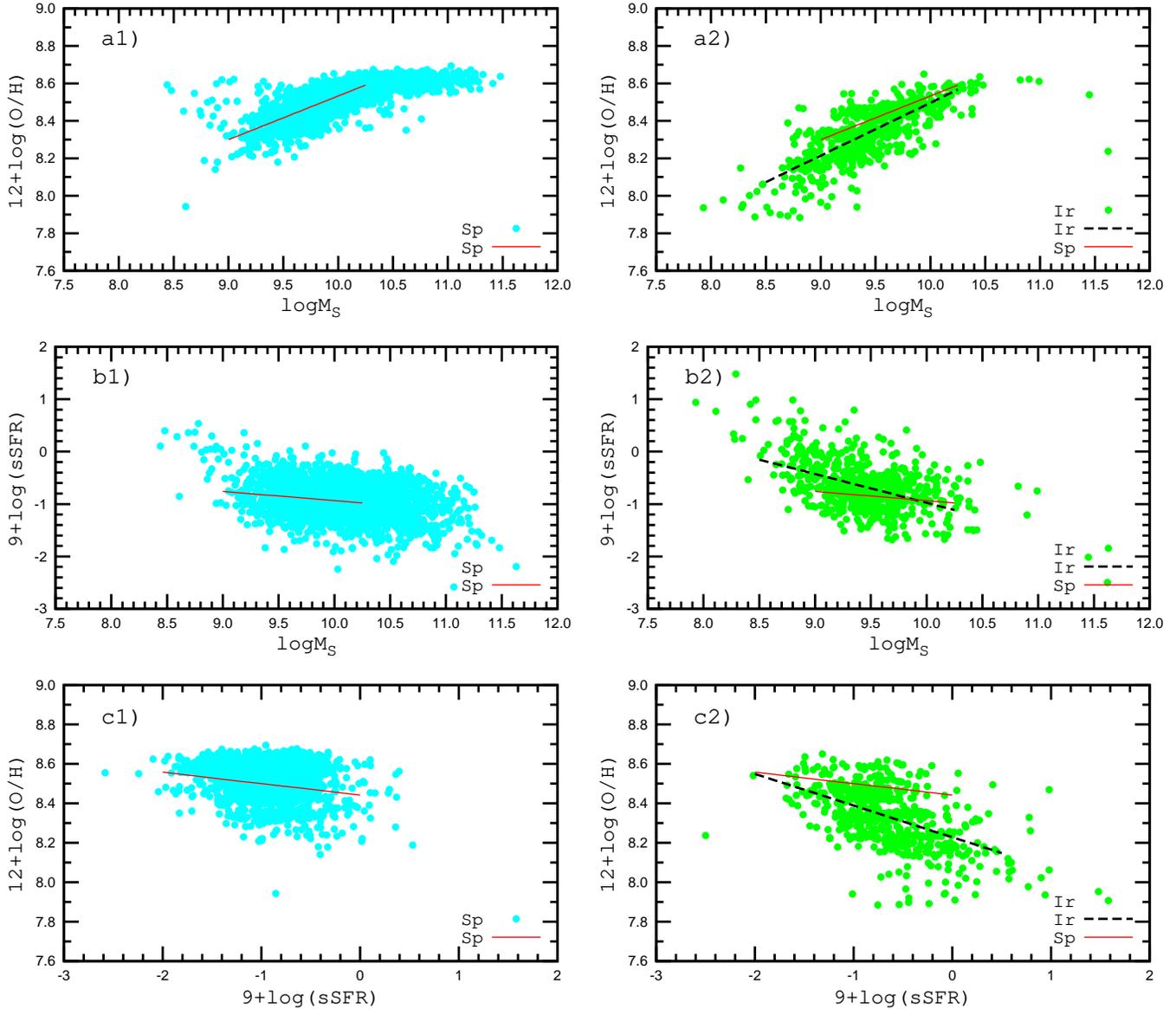}}
\caption{ 
{\it Panel a1).} 
The mass -- metallicity diagram for spiral galaxies. 
Light-gray (blue) points show individual objects, while
the dark-gray (red) solid line shows the best fit to galaxies 
with the masses between log$M_S$ = 9.0 and log$M_S$ = 10.25.
{\it Panel a2).} 
The mass -- metallicity relation for irregular galaxies. 
Light-gray (green) points are individual objects, while 
the dark (black) dashed line is the best fit to galaxies 
with masses between log$M_S$ = 8.5 and log$M_S$ = 10.25.
The dark-gray (red) solid line is the same as in panel $a1)$.
{\it Panels b1) and b2)} show the  mass -- $sSFR$ diagrams for the 
same samples of spiral and irregular galaxies, respectively. 
{\it Panels c1) and c2)} show the  $sSFR$ -- metallicity diagrams for 
spiral and irregular galaxies, respectively. 
(A colour version of this figure is available in the online version.) 
}
\label{figure:ms-oh}
\end{figure*}

Thus, there is no correlation between oxygen abundance and $sSFR$ in massive galaxies
and the  oxygen abundance correlates with the $sSFR$ in low-mass galaxies.  
The difference in behaviour of massive and low-mass galaxies may be caused by the 
fact that our sample of low-mass galaxies is a mixture of spiral and irregular
galaxies while  massive galaxies can be expected to 
constitute subsamples of galaxies of a single morphological type (spiral galaxies).  
Below, we compare the properties of irregular and spiral galaxies of low masses.

\subsection{Spirals versus irregulars}

Let us firstly test whether the spiral and irregular galaxies of similar masses from 
our sample have different oxygen abundances. 
Morphological classifications of the SDSS galaxies have been performed within 
the framework of "Galaxy Zoo" project\footnote{The catalogs are available at 
http://www.galaxyzoo.org}. The data catalogue are available in SDSS DR9 database.
The project is described in \citet{Lintottetal2008} and the data release is 
described in \citet{Lintottetal2011}. The fraction of the votes in each of the six categories 
(elliptical, clockwise spiral, anticlockwise spiral, edge-on disc, don't know, merger) 
is given, along with "debiased" votes in elliptical and spiral categories
and flags identifying systems as classified as spiral, elliptical or uncertain.
We assume the votes of spiral galaxies is the sum of the fraction of the votes of 
clockwise spiral and counterclockwise spiral. Thus, each galaxy is described 
by five values $f_X$ which specify the probability that the galaxy belongs to one 
of those classes. We assume that the galaxy is a spiral galaxy if the value 
of $f_{Sp}$ $\ga$ 2/3. Since the class "irregular galaxies" is not included,
we select irregular galaxies by elimination, i.e., we assume that 
the galaxy is an irregular galaxy if each  value 
$f_{Sp}$, $f_{E}$, $f_{Edge}$, $f_{Merg}$ is less than 1/3.

If there is a difference between oxygen abundances in irregular and 
spiral galaxies of similar masses then 
one may expect the mass -- metallicity relation for irregular galaxies 
to be different from that for spiral galaxies. 
The panel $a1)$ of  Fig.~\ref{figure:ms-oh} shows the mass -- metallicity diagram for 
spiral galaxies at the present epoch (for 
galaxies with redshifts less than 0.05). Light-gray (blue) points are individual
galaxies. Oxygen abundances in massive galaxies 
do not show significant correlation with galaxy mass. This flattening of the mass -- 
metallicity relation (a plateau at high masses) has previously been observed by 
\citet{Tremontietal2004,Thuanetal2010}. The luminosity -- metallicity relation 
for nearby spiral galaxies also exhibits a plateau at high luminosities \citep{Pilyuginetal2007MNRAS376}.
Since our goal is to compare the mass -- metallicity relations for spiral and irregular 
galaxies we should derive the relations for a galaxy mass range well populated with
both spiral and irregular galaxies, log$M_S$ $\la$ 10.25. 
The best linear least-squares fit to the data for spiral galaxies with masses 10.25 $\ga$ log$M_S$ $\ga$ 9.0 is
\begin{equation}
12 + \log ({\rm O/H}) = 6.20 (\pm 0.05) + 0.233 (\pm 0.006) \times \log M_S  , 
\label{equation:sp}   
\end{equation}
shown in the panel $a1)$  of  Fig.~\ref{figure:ms-oh} as the dark-gray (red) 
solid line. The small formal errors in the coefficients of Eq.(\ref{equation:sp}) 
and the rather small mean deviation (0.057 dex) are due to the relatively large number of 
data points (1494).

The panel $a2)$  of  Fig.~\ref{figure:ms-oh} shows the mass -- metallicity diagram for 
irregular galaxies at the present epoch, $z$ $\la$ 0.05. Light-gray (green) 
points are individual galaxies. The linear best fit to the data for irregular  
galaxies with masses 10.25 $\ga$ log$M_S$ $\ga$ 8.5 
\begin{equation}
12 + \log ({\rm O/H}) = 5.68 (\pm 0.10) - 0.282 (\pm 0.011) \times \log M_S  , 
\label{equation:ir}   
\end{equation}
is presented in the panel $a2)$  of  Fig.~\ref{figure:ms-oh} by the dark (black) 
dashed line, while the dark-gray (red) solid line in the lower panel is the same as in 
the panel $a1)$. Comparison of these slopes, as well as comparison 
between  Eq.(\ref{equation:sp}) and Eq.(\ref{equation:ir}), shows that 
the mass -- metallicity relation for irregular galaxies differs slightly
from that for spiral galaxies. 
Note in passing, however, that the difference is rather small, less than the scatter 
in abundances in both irregular and spiral galaxies. 
 
Panel $b1)$ of  Fig.~\ref{figure:ms-oh} shows the mass -- $sSFR$ diagram for 
spiral galaxies with redshifts less than 0.05. Light-gray (blue) points are individual
galaxies.  The best linear least-squares fit to the data for spiral galaxies 
with masses 10.25 $\ga$ log$M_S$ $\ga$ 9.0 is
\begin{equation}
9 + \log (sSFR) = 0.81 (\pm 0.28) - 0.17 (\pm 0.03) \times \log M_S  , 
\label{equation:spb}   
\end{equation}
shown in panel $b1)$  of  Fig.~\ref{figure:ms-oh} as the dark-gray (red) solid line. 
Panel $b2)$  of  Fig.~\ref{figure:ms-oh} shows the mass --  $sSFR$ diagram for 
irregular galaxies.  The linear best fit to the data for irregular  
galaxies with masses 10.25 $\ga$ log$M_S$ $\ga$ 8.5 
\begin{equation}
9 + \log (sSFR) = 4.51 (\pm 0.43) - 0.55 (\pm 0.05) \times \log M_S  , 
\label{equation:irb}   
\end{equation}
is presented in panel $b2)$  of  Fig.~\ref{figure:ms-oh} as a dark (black) 
dashed line, while the dark-gray (red) solid line in the lower panel is the same as in 
panel $b1)$. 
Comparison of  Eq.(\ref{equation:spb}) and Eq.(\ref{equation:irb}), 
shows that the mass -- $sSFR$  relation for irregular galaxies is steeper than 
that for spiral galaxies. Spiral and irregular galaxies with masses  log$M_S$ $\sim$ 10 
have similar $sSFR$s, while 
irregular galaxies with masses  log$M_S$ $\sim$ 9 have higher $sSFR$s than spiral 
galaxies of similar masses. 

Panel $c1)$ of  Fig.~\ref{figure:ms-oh} shows the $sSFR$ -- metallicity diagram for 
spiral galaxies with redshifts less than 0.05. Light-gray (blue) points are individual
galaxies.  The best linear least-squares fit to the data for spiral galaxies 
with masses 10.25 $\ga$ log$M_S$ $\ga$ 9.0 is
\begin{equation}
12 + \log ({\rm O/H}) = 8.44 (\pm 0.01) - 0.058 (\pm 0.007) \times (\log (sSFR) + 9)   , 
\label{equation:spc}   
\end{equation}
shown in panel $c1)$  of  Fig.~\ref{figure:ms-oh} as a dark-gray (red) solid line. 
Panel $c2)$  of  Fig.~\ref{figure:ms-oh} shows the  $sSFR$ -- metallicity diagram for 
irregular galaxies.  The linear best fit to the data for irregular  
galaxies with masses 10.25 $\ga$ log$M_S$ $\ga$ 8.5 
\begin{equation}
12 + \log ({\rm O/H}) = 8.23 (\pm 0.01) - 0.160 (\pm 0.011) \times (\log (sSFR) + 9)   , 
\label{equation:irc}   
\end{equation}
is presented in panel $c2)$  of  Fig.~\ref{figure:ms-oh} as a dark (black) 
dashed line, while the dark-gray (red) solid line in the lower panel is the same as in 
the panel $c1)$. 
Comparison of  Eq.(\ref{equation:spc}) and Eq.(\ref{equation:irc}), 
shows that the $sSFR$ -- metallicity  relation for irregular galaxies is steeper than 
that for spiral galaxies. 
 
The fact that our sample of low-mass galaxies is a mixture of spiral and irregular
galaxies may explain the dependence of the  metallicity -- redshift relation 
on the $sSFR$ observed for low-mass SDSS galaxies.
The subsample of low-mass galaxies with star formation rates close to $sSFR_{max}$ contains mainly 
irregular galaxies while 
the subsample of low-mass galaxies with star formation rates around  0.1$\times$$sSFR_{max}$  
contains both irregular and spiral galaxies. 
Therefore these subsamples show different properties and the metallicity -- redshift relations 
for low-mass galaxies with $sSFR$ = $sSFR_{max}$ and $sSFR$ =  0.1$\times$$sSFR_{max}$ 
are different. 

It is expected that in general the oxygen abundance in a galaxy 
correlates with three parameters, i.e., O/H = $f$($M_S$, $sSFR$, $z$).
Using the same data points as in the case of Eq.(\ref{equation:sp}), we have found for 
spiral galaxies
\begin{eqnarray}
       \begin{array}{lll}
12 + \log ({\rm O/H})   & =  &   6.18 \, (\pm 0.05) + 0.240 \, (\pm 0.006) \;  \log M_S   \\
      & -  &  0.023 \, (\pm 0.005)\; (\log sSFR + 9)                                 \\
      & -  &  1.97\, (\pm 0.22) \; z  .                                               \\
     \end{array}
\label{equation:sp1}
\end{eqnarray}
Again, using the same data points as in the case of Eq.(\ref{equation:ir}), we have found for 
irregular galaxies
\begin{eqnarray}
       \begin{array}{lll}
12 + \log ({\rm O/H})   & =  &   5.90 (\pm 0.11) + 0.262 (\pm 0.001) \log M_S   \\
      & -  &  0.069 (\pm 0.009) (\log sSFR + 9)                                 \\
      & -  &  2.13 (\pm 0.56) z                        .                         \\
     \end{array}
\label{equation:ir1}
\end{eqnarray}
Comparison of the coefficients in Eq.(\ref{equation:sp1}) and Eq.(\ref{equation:ir1}) shows 
that the largest difference (by a factor of $\sim$3) is found for the coefficients in the term with  $sSFR$.

\subsection{Comparison of the mass -- metallicity and luminosity -- metallicity relations 
for SDSS and nearby irregulars}

\subsubsection{The mass -- metallicity relations}

\begin{figure}
\resizebox{1.00\hsize}{!}{\includegraphics[angle=000]{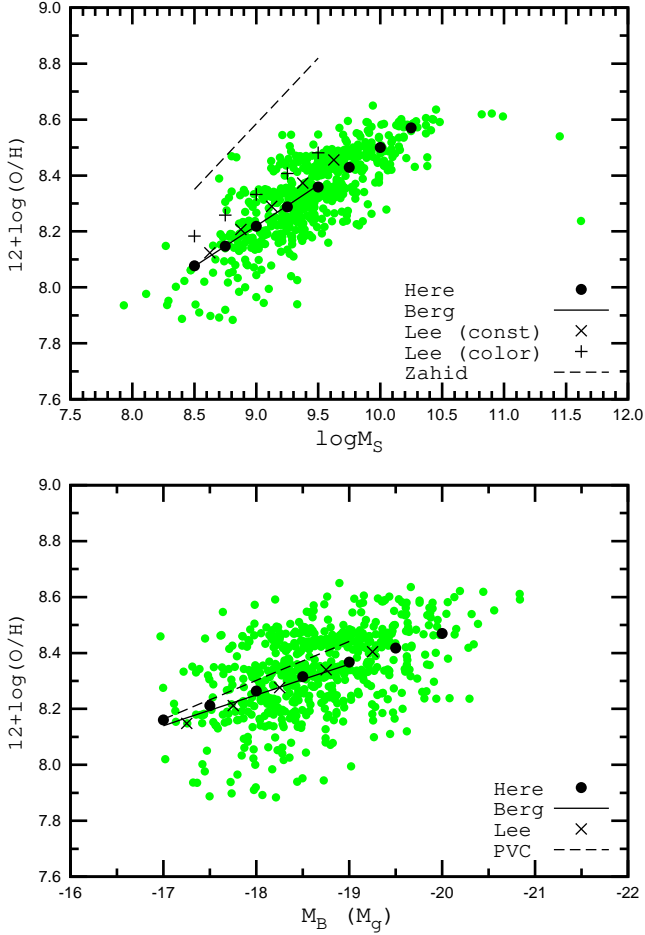}}
\caption{
{\it Upper panel.} 
The mass -- metallicity relation for irregular galaxies. 
The gray (green) points show SDSS galaxies (same as in Fig.~\ref{figure:ms-oh}),
while dark (black) points show the best fit to data for stellar masses between log$M_S$ = 8.5 and  log$M_S$ = 10.25.  
The dark (black) solid line is the relation for nearby irregular galaxies from \citet{Bergetal2012} and
the crosses show the mass -- metallicity  relation for nearby irregulars 
derived by \citet{Leeetal2006ApJ647} with constant mass-to-light ratios, 
while the plus signs show their relation for $M_S$ obtained with colour-dependent mass-to-light ratios.
The dark (black) dashed line is the mass -- metallicity  relation for 
SDSS galaxies from  \citet{Zahidetal2012ApJ750}. 
{\it Lower panel.} 
The luminosity -- metallicity relation for irregular galaxies. 
The  gray (green) points are the SDSS galaxies, while
the dark (black) points show the best fit to data for luminosities between $M_g$ = -20 and  $M_g$ = -17.
The dark (black) solid line is the relation for nearby irregular galaxies from \citet{Bergetal2012} and
the crosses show the luminosity -- metallicity  relation for nearby irregulars 
according to \citet{Leeetal2006ApJ647}. 
The dark (black) dashed line is the $M_B$ -- (O/H)$_{T_{e}}$ relation for 
nearby irregulars from \citet{Pilyuginetal2004AA425}. 
(A colour version of this figure is available in the online version.) 
}
\label{figure:lz}
\end{figure}

Fig.~\ref{figure:lz} shows a comparison between our mass -- metallicity and luminosity -- metallicity   
relations for SDSS galaxies and relations obtained for nearby irregular galaxies. 
The upper panel shows the mass -- metallicity diagram, where
the gray (green) points are irregular galaxies from the SDSS, the same as in Fig.~\ref{figure:ms-oh}. 
The dark (black) points show the best fit to the data in the mass range
from log$M_S$ = 8.5 to  log$M_S$ = 10.25.  
\citet{Leeetal2006ApJ647} have constructed mass -- metallicity and luminosity -- metallicity 
relations for 25 nearby irregular galaxies. The oxygen abundances in 21 of these galaxies were derived 
through the $T_e$ method. They estimated the masses of galaxies from their luminosities in two ways. 
In the first case they adopted a constant stellar-mass-to-NIR-light ratio and obtained 
the following mass -- metallicity relation,
\begin{equation}
12 + \log ({\rm O/H}) = 5.26 (\pm 0.27) + 0.332 (\pm 0.033) \times \log M_S.
\label{equation:msleeconst}
\end{equation}
This relation is shown in the upper panel of Fig.~\ref{figure:lz} by dark (black) crosses. 
In the second case they adopted a colour-dependent stellar-mass-to-NIR-light ratio, which yielded
the following mass -- metallicity relation,
\begin{equation}
12 + \log ({\rm O/H}) = 5.65 (\pm 0.23) + 0.298 (\pm 0.030) \times \log M_S.
\label{equation:msleecolour}
\end{equation}
This relation is shown in the upper panel of Fig.~\ref{figure:lz} by dark (black) plus signs.
Furthermore, \citet{Bergetal2012} measured the temperature sensitive [O\,{\sc iii}]$\lambda$4363 line for 
31 low luminosity galaxies in the {\it Spitzer} Local Volume Legacy survey, and thus determined 
oxygen abundances using the direct method. They have analysed a "Combined Select" sample composed of 38 objects 
(taken from their sample and as well as the literature) with direct oxygen abundances and reliable distance 
determinations. Stellar masses where estimated from the 4.5 $\mu$m luminosities and 
$K$ -- [4.5] and $B$ -- $K$ colours.  Using these data, they derived a mass -- metallicity relation, 
\begin{equation}
12 + \log ({\rm O/H}) = 5.61 (\pm 0.24) + 0.29 (\pm 0.03) \times \log M_S .
\label{equation:msberg}
\end{equation}
This relation is shown as the dark (black) solid line 
in the upper panel of Fig.~\ref{figure:lz}.

The upper panel of Fig.~\ref{figure:lz} shows that our mass -- metallicity relation, 
(O/H)$_{C_{ON}}$ -- $M_S$, for SDSS galaxies are consistent with (O/H)$_{T_e}$ -- $M_S$ relations
for nearby irregular galaxies obtained by \citet{Leeetal2006ApJ647} and \citet{Bergetal2012}. 

\citet{Zahidetal2012ApJ750} have estimated oxygen abundances in the SDSS galaxies using the 
empirical $N2$ calibration after \citet{Yinetal2007AA462} and have constructed a mass -- metallicity relation, 
\begin{equation}
12 + \log ({\rm O/H}) = 5.585 (\pm 0.003) + 0.47 (\pm 0.01) \times (\log M_S - 9) 
\label{equation:zahid}
\end{equation}
which is shown in the upper panel of Fig.~\ref{figure:lz} by dark (black) dashed line.
This mass -- metallicity relation for SDSS galaxies differs 
significantly from the mass -- metallicity relation for SDSS galaxies considered here
and from the the mass -- metallicity relations for nearby galaxies by 
\citet{Leeetal2006ApJ647} and \citet{Bergetal2012}.
This may appear surprising, since 
\citet{Zahidetal2012ApJ750} have estimated oxygen abundances using the purely empirical $N2$ calibration. 
However,  \citet{Yinetal2007AA462} have noted that the $N2$ index is only a useful metallicity indicator for 
galaxies with 12 + log(O/H) $\la$ 8.5 as they had no calibrating data 
points with 12 + log(O/H) $\ga$ 8.5. Hence, the high oxygen abundances (up to 12 + log(O/H) $\sim$ 9.0) 
obtained by \citet{Zahidetal2012ApJ750} with this calibration may be disputed.

\subsubsection{The luminosity -- metallicity relations}

The lower panel  of Fig.~\ref{figure:lz} shows the luminosity -- metallicity diagram
for irregular galaxies. 
The  gray (green) points show O/H -- $M_g$ diagram for irregular galaxies 
from our SDSS sample, 
where $M_g$ is the absolute magnitude of a galaxy in the SDSS photometric band $g$. 
The dark (black) points show the best fit to those data, 
\begin{equation}
12 + \log ({\rm O/H}) = 6.41 (\pm 0.16) - 0.103 (\pm 0.009) \times  M_g,
\label{equation:mbour}
\end{equation}
obtained in the luminosity range from $M_g$ = -20 to  $M_g$ = -17.  
The $B$ -- $g$ colour index is $\sim 0.1$ mag only for the star-forming galaxies 
and the O/H -- $M_g$ and  O/H -- $M_B$ relations are therefore comparable
\citep{Papaderosetal2008AA491,Gusevaetal2009}.
The O/H -- $M_B$ relation obtained by \citet{Bergetal2012} 
for their "Combined Select" sample of nearby irregular galaxies,
\begin{equation}
12 + \log ({\rm O/H}) = 6.27 (\pm 0.21) - 0.11 (\pm 0.01) \times  M_B,
\label{equation:mbberg}
\end{equation}
is shown by the dark (black) solid line. 
It should be noted that this relation seems to be valid up to $M_B$ $\sim$ --8 
\citep{Skillmanetal2013}.
The relation for nearby irregular galaxies by \citet{Leeetal2006ApJ647},
\begin{equation}
12 + \log ({\rm O/H}) = 5.94 (\pm 0.27) - 0.128 (\pm 0.017) \times  M_B,
\label{equation:mblee}
\end{equation}
is shown by dark (black) crosses, and that by \citet{Pilyuginetal2004AA425},
\begin{equation}
12 + \log ({\rm O/H}) = 5.80 (\pm 0.17) - 0.139 (\pm 0.011) \times  M_B,
\label{equation:mbpvc}
\end{equation}
is represented by the dark (black) dashed line.  

Our luminosity -- metallicity 
relation, (O/H)$_{C_{ON}}$ -- $M_g$, for SDSS galaxies is in satisfactory agreement with (O/H)$_{T_e}$ -- $M_B$ 
relations for nearby irregular galaxies obtained in several other studies \citep{Bergetal2012,Leeetal2006ApJ647,Pilyuginetal2004AA425}. 
It should also be noted that the galaxy metallicity correlates more tightly with its stellar mass than
with its luminosity.

\section{Discussion}

\subsection{Does the aperture effect contribute to the redshift dependence of the oxygen 
abundances?}

\begin{figure*}
\resizebox{1.00\hsize}{!}{\includegraphics[angle=000]{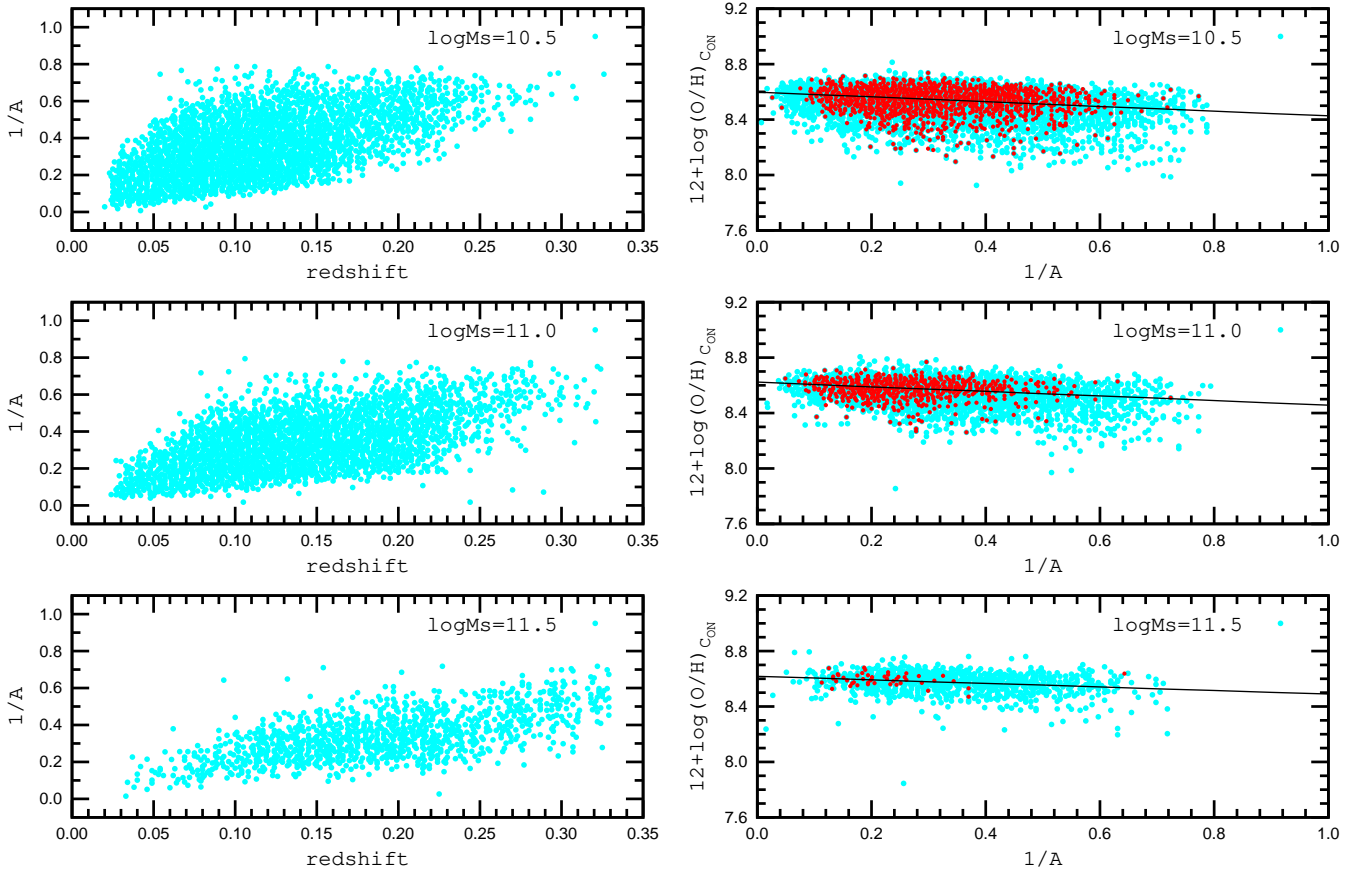}}
\caption{ 
{\it Left panels.} 
Redshift dependence of the galaxy light fraction within the fiber 
(1/$A$) for galaxies of different masses.
{\it Right panels.} 
The oxygen abundance in a galaxy as a function of 1/$A$. 
Light gray (blue) points are individual objects with redshifts 
from the full considered range, 0.02 $\la$ $z$ $\la$ 0.33. 
Dark (black) line  is the best fit to  those data. 
Dark gray (red) points are individual objects with redshifts from 
the range 0.08 -- 0.1. 
(A colour version of this figure is available in the online version.) 
}
\label{figure:z-a}
\end{figure*}

The right panels in  Fig.~\ref{figure:z-sfr} show there is 
a redshift dependence of the oxygen abundance in galaxies of different masses
in the sense that the oxygen abundances in distant galaxies are lower than 
that in nearby galaxies. Chemical evolution of galaxies is usually considered 
the only reason for the redshift variation of the oxygen abundance 
in galaxies. 
In the case of the SDSS galaxies there seems to be another effect that could make a
contribution to this redshift dependence.
The SDSS galaxy spectra span a relatively wide range of redshifts. 
There is thus an aperture-redshift effect present in the
SDSS data (all spectra are obtained with 3-arcsec-diameter fibers). 
At a low redshift the projected aperture diameter is smaller than 
that at high redshift. 
This means that, at low redshifts,  SDSS spectra can be seen as global spectra of the
central parts of galaxies (if the fibers are centered on centers of galaxies), 
i.e., composite nebulae including  H\,{\sc ii} regions near the center.
At large redshifts, however, SDSS spectra are closer to global spectra of 
whole galaxies, i.e., spectra of  composite nebulae including  H\,{\sc ii} regions 
near the center as well as in the periphery. 
Since there is usually a radial abundance gradient in the discs of spiral galaxies 
\citep[][among others]{Vilacostas1992,Zaritskyetal1994,Pilyuginetal2004AA425},  
the oxygen abundance in the composite nebulae in distant galaxies may be lower 
than that in nearby galaxies if the fibers are centered on centers of galaxies.

We estimate the aperture-correction factor $A$ (see Eq.\ref{equation:a}) to account for the fact 
that only a limited amount of emission from a galaxy is detected through the 3" diameter fiber.  
The quantity 1/$A$ is the galaxy light fraction within the fiber. 
The left panels of Fig.~\ref{figure:z-a} show the redshift dependence of the 
1/$A$ for galaxies of different masses.
The right panels of Fig.~\ref{figure:z-a} show the oxygen abundance in a galaxy 
as a function of 1/$A$  for subsamples of galaxies of different masses. 
Light gray (blue) points a show individual objects with redshifts 
in the full considered range, 0.02 $\la$ $z$ $\la$ 0.33 and the dark (black) line  is the best fit to  the data. 
Dark gray (red) points show individual objects with redshifts in
the range 0.08 -- 0.1. 
There is obviously a correlation between the oxygen abundance in a galaxy and the galaxy 
light fraction within the fiber. 
The correlations for the full considered redshift range 
and that for a restricted redshift range are similar.
The galaxy light fraction within the fiber also correlates with 
the redshift. Therefore, the diagrams in Fig.~\ref{figure:z-a} cannot tell us
whether the correlation between the oxygen abundance in a galaxy and 
the galaxy light fraction within the fiber is caused by redshift evolution 
(enrichment in heavy elements) of galaxies or by the aperture effect or a combination of both. 

\begin{figure}
\resizebox{1.00\hsize}{!}{\includegraphics[angle=000]{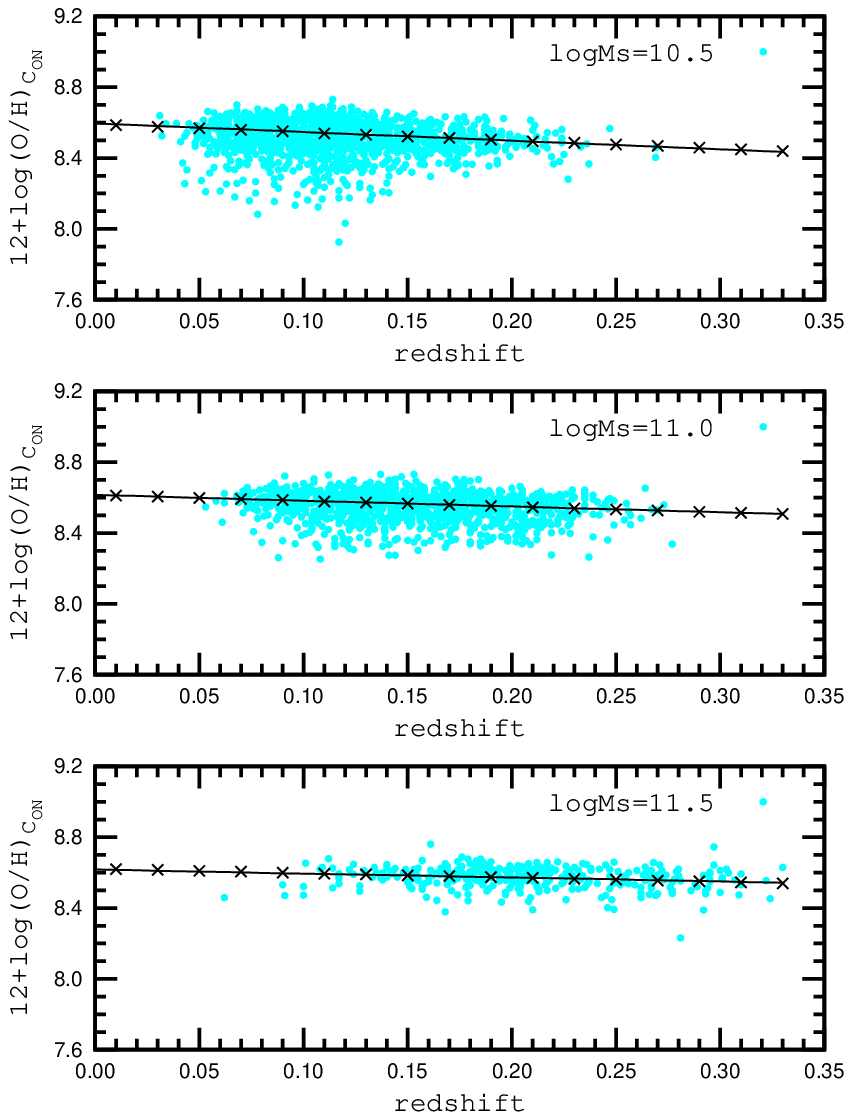}}
\caption{
Redshift dependence of the oxygen abundances in galaxies of different masses 
and an aperture correction factor  0.45 $>$ 1/$A$  $>$ 0.35. 
Gray (blue) points are individual objects, while the dark (black) crosses show 
the best fit to the data. 
The dark (black) line  is the best fit to all galaxies in the
given mass range.
(A colour version of this figure is available in the online version.) 
}
\label{figure:z-oh-a04}
\end{figure}

To solve this dilemma we compare the redshift variation of the oxygen 
abundance in subsample of galaxies for a given value of the aperture correction factor
and that for total sample of galaxies of a given mass.
Gray (blue) points in the upper panel of Fig.~\ref{figure:z-oh-a04} show 
the redshift dependence of the oxygen abundance with galaxy masses log$M_S$ 
ranging from 10.25 to 10.75, and 1/$A$ in the  range 0.45 $>$ 1/$A$  $>$ 0.35. 
The dark (black) crosses is the linear best fit to those data 
\begin{equation}
12 + \log ({\rm O/H}) = a + b \times z .
\label{equation:ohz}   
\end{equation}
Values for the coefficients $a$ and $b$ of Eq.(\ref{equation:ohz}) are listed in Table~\ref{table:ohz}. 
The small statistical uncertainties in the coefficients $a$ and $b$ 
is due to the large numbers of galaxies in the subsamples. 
The dark (black) line is the linear best fit to all galaxies within
the mass range considered, which essentially coincide with the crosses representing the subsample.

Similar diagrams for the other mass intervals 
are presented in the middle and lower panels of Fig.~\ref{figure:z-oh-a04}. 
Again, the relation O/H -- $z$ for the subsample and the total sample of 
galaxies within the given mass ranges coincide.
This means that the aperture effect does not 
make any significant contribution to the redshift dependence of the oxygen 
abundance in galaxies. 
The trend in oxygen abundance with redshift must be caused by the chemical evolution of galaxies, 
i.e. by the increasing level of astration at low redshifts.

\begin{table}
\caption[]{\label{table:ohz}
Coefficients in the "oxygen abundance -- redshift" relations, 
12 + log(O/H) = $a$ + $b$$\times$$z$, for subsamples of galaxies of different 
stellar masses and different fraction of the light within fiber  1/$A$. 
(A colour version of this figure is available in the online version.) 
}
\begin{center}
\begin{tabular}{cccc} \hline \hline
mass range    &   range of 1/$A$          &   $a$             &    $b$           \\ 
log$M_S$      &                           &                   &                   \\ \hline
10.25 - 10.75 & 0.45 $>$ 1/$A$  $>$ 0.35  &  8.591$\pm$0.004  &  -0.456$\pm$0.037 \\ 
              & 1    $>$ 1/$A$  $>$ 0     &  8.596$\pm$0.001  &  -0.488$\pm$0.012 \\ 
              &                           &                   &                   \\ 
10.75 - 11.25 & 0.45 $>$ 1/$A$  $>$ 0.35  &  8.615$\pm$0.005  &  -0.326$\pm$0.034 \\ 
              & 1    $>$ 1/$A$  $>$ 0     &  8.615$\pm$0.002  &  -0.325$\pm$0.012 \\ 
              &                           &                   &                   \\ 
11.25 - 11.75 & 0.45 $>$ 1/$A$  $>$ 0.35  &  8.623$\pm$0.012  &  -0.250$\pm$0.057 \\ 
              & 1    $>$ 1/$A$  $>$ 0     &  8.617$\pm$0.004  &  -0.222$\pm$0.022 \\ 
\hline
\end{tabular}\\
\end{center}
\end{table}

The fact that the aperture effect does not play any significant role 
in the redshift dependence of the oxygen abundance may be 
explained if the fiber positions are distributed 
randomly over the images of galaxies. If this is the case, 
even in nearby galaxies, the fibers may cover 
both the central and the periphery parts of the galaxies. In such case, 
the average oxygen abundances in a subsample of SDSS 
galaxies is more similar to the characteristic oxygen abundance in a galaxy rather than 
to its central oxygen abundance.  
It should be noted that many observations with the fibers centered on 
the galaxy nuclei are excluded from our analysis since the spectra are often 
classified as AGNs and not star-forming regions.

\subsection{Time-averaged specific star formation rate in massive galaxies}

The trend in oxygen abundance with redshift is caused by the increasing level 
of astration (or by the decreasing of the gas-mass fraction $\mu$ = $M_{gas}$/($M_{S}$+$M_{gas}$))  
with decreasing of redshift. 
This provides a possibility to estimate the value of the star formation rate 
in galaxies of different masses averaged over a fixed time interval. 
We will consider here only massive galaxies, which can be expected to 
constitute subsamples of galaxies of a single morphological type (spiral galaxies).  

The prediction of the simple (closed-box) model of galactic chemical evolution
can be well approximated by a linear expression for $\mu$ values
between $\sim$ 0.7 and $\sim$ 0.1. 
In a real situation, the oxygen abundance is also affected by the mass exchange
between a galaxy and its environment \citep{pagel1997}.
It is commonly accepted that gas infall plays an important
role in the chemical evolution of disks of spiral galaxies. Therefore, the
application of the simple model to large spiral galaxies may appear unjustified. 
It is expected that the rate of gas infall onto the disk decreases exponentially with time
\citep[e.g.]{Caluraetal2009AA504}. 
It has been shown \citep{PilyuginFerrini1998AA336}
that the present-day location of a system in the $\mu$ -- O/H
diagram is governed mainly by its evolution in the recent past, but is
weakly dependent of its evolution on long timescales.
Therefore, the fact that the
present-day location of spiral galaxies is near that predicted by
the simple model is not in conflict with infall of
gas onto the disk taking place over
a long time (the latter is assumed necessary to satisfy the
observed abundance distribution function and the age -- metallicity relation
in the solar neighbourhood) since these observational data reflect the
evolution of the system in a distant past.
A decrease of $\mu$ by 0.1 results in a 
increase of oxygen abundance by $\sim$ 0.13 dex \citep{Pilyuginetal2006MNRAS367}, 
which lead us to the approximation
\begin{equation}
 \Delta (\log({\rm O/H})) \approx -1.3\times\Delta\mu.
\label{equation:ohmu}   
\end{equation}
The variation in oxygen abundance with redshift is given by the cooefficient $b$ in 
 Eq. (\ref{equation:ohz}), the value of $-0.1b$ corresponds to a change in oxygen abundance 
$\Delta$(log(O/H)) over a redshift interval $\Delta$$z$ = 0.1. At low redshifts, 
a redshift interval $\Delta$$z$ = 0.1 corresponds to a time interval of about 1 Gyr. 
Taking Eq.(\ref{equation:ohmu}) into account, the change of gas-mass fraction 
$\Delta$$\mu$  over the time interval of 1 Gyr corresponds to 0.077$b$.
The gas mass fraction is expressed in terms of the total mass of a galaxy $M_{tot}$. 
The specific star formation rate  is expressed in terms of the stellar mass of a 
galaxy $M_S$. The oxygen abundance 12 + log(O/H) = 8.6 (which is a typical value 
for massive galaxies,  Fig.~\ref{figure:z-oh-a04}) corresponds to the value of 
gas mass fraction $\mu$ $\sim$ 0.2 \citep{Pilyuginetal2006MNRAS367}. 
This gives $M_S$ $\approx$ 0.8$\times$$M_{tot}$.
Then the change of gas mass per year expressed in terms of the stellar mass of a 
galaxy (i.e. the average $sSFR$) is given by the value 10$^{-10}$$\times$$b$.

Using the values of $b$ from the Table~\ref{table:ohz}, we estimate the 
time-averaged $sSFR$s for massive galaxies. We find 
9+log($sSFR_{aver}$) $\sim$ -1.31 for galaxies of masses log$M_S$ = 10.5 +/-0.2, 
9+log($sSFR_{aver}$) $\sim$ -1.49 for galaxies of masses log$M_S$ = 11.0 +/-0.2, 
and 9+log($sSFR_{aver}$) $\sim$ -1.64 for galaxies of masses log$M_S$ = 11.5 +/-0.2. 
Comparison between the $sSFR_{max}$ (Fig.~\ref{figure:z-sfr}) and $sSFR_{aver}$ 
shows that the  $sSFR_{max}$ exceeds the  $sSFR_{aver}$ by a factor of 6-7. 
Since the $sSFR_{max}$ we obtain is not 
an absolute maximum star formation rate but the mean value for only
5-10\% of galaxies with highest $sSFR$s (see Fig.~\ref{figure:z-sfr}),
the absolute maximum star formation rate can exceed the time-averaged 
star formation rate in galaxies of a given mass by about an order of magnitude. 

\section{Summary and conclusions}

In the present study, we have used a sample of 98.986 star-forming SDSS galaxies to examine the dependence 
of the metallicity -- redshift relations on the star formation rate for galaxies 
of different stellar masses. 
We determined two values of oxygen and nitrogen abundances for each object 
using a modified version of the recent counterpart method ($C$ method) 
\citep{Pilyuginetal2012MNRAS424}. 
The (O/H)$_{C_{\rm ON}}$ and (N/H)$_{C_{\rm ON}}$ abundances were derived using the 
log$R_3$,  $P$ and  log($N_2$/$R_2$) values to find a counterpart for the studied SDSS galaxy,
and, similarly, the (O/H)$_{C_{\rm NS}}$ and (N/H)$_{C_{\rm NS}}$ abundances were derived using the
log$R_3$, log$N_2$ and log($N_2$/$S_2$) values to find a counterpart.  
In total, we used a sample of 233 reference H\,{\sc ii} regions for abundance determinations. 
The conditions $|$log(O/H)$_{C_{NS}}$ -- log(O/H)$_{C_{ON}}$$|$ $\le$ 0.05 and 
$|$log(N/H)$_{C_{NS}}$ -- log(N/H)$_{C_{ON}}$$|$ $\le$ 0.10 were used to select the 
SDSS objects with the most reliable abundances. 
Only the (O/H)$_{C_{\rm ON}}$ abundances were used in our further analyses. 

We analysed the redshift dependence of the maximum specific star formation 
rate $sSFR_{max}$ in SDSS galaxies of different masses. 
Six stellar mass intervals centered on the values log$M_s$ = 9.0, 
9.5, 10.0, 10.5, 11.0, 11.5 and width interval widths +/- 0.2 dex were considered. 
We find that the $sSFR_{max}$ decreases with increasing stellar mass 
and decreasing redshift. 
For each mass interval we considered the metallicity -- redshift relations for 
subsamples of galaxies with $sSFR$ $\sim$ $sSFR_{max}$ and with $sSFR$ $\sim$ 0.1$\times$$sSFR_{max}$.
These relations coincide for massive (log$M_S$ $\ga$ 10.5) galaxies, i.e., 
the metallicity -- redshift relations for massive galaxies are independent of the $sSFR$.
On the other hand, these relations differ for low-mass galaxies as the 
galaxies with $sSFR$ $\sim$ $sSFR_{max}$ have, on average, lower oxygen abundances than 
galaxies with $sSFR$ $\sim$ 0.1$\times$$sSFR_{max}$.

We also find evidence in favour of irregular galaxies 
having higher (on average) $sSFR$s and lower oxygen abundances than spiral 
galaxies of similar masses and that the mass -- metallicity relation 
for spiral galaxies differs slightly from that for irregular galaxies.
The mass -- metallicity and luminosity -- metallicity relations obtained for 
irregular SDSS galaxies agree with the same for nearby irregular galaxies 
with direct abundance determinations. 

The fact that any sample of low-mass galaxies is likely a mixture of spiral and 
irregular galaxies can explain the dependence of the  metallicity -- redshift 
relation on the $sSFR$.

The redshift variation of the oxygen abundances in galaxies of a given mass 
allows us to estimate the decrease of gas-mass fractions, i.e. 
this provides a possibility to estimate the star formation rate 
in galaxies of different masses averaged over a given time interval. 
We estimated the time-average specific star formation rates for massive galaxies. 
We find that the absolute maximum star formation rate can exceed the time-averaged 
star formation rate in massive galaxies by about an order of magnitude. 

Finally, we compared  the relations O/H -- $z$ for subsample of galaxies within a given
range of galaxy light fractions within the spectroscopic fiber and for 
total sample of galaxies of a given mass.
These subsamples show very similar relations, which
suggests the aperture effect does not make a significant contribution to 
the redshift dependence of the oxygen abundances in galaxies. 
This may be explained by fiber positions being distributed 
randomly over the images of the galaxies, so in galaxies the fibers may cover 
both central and periphery parts of different galaxies.In such case, 
the average oxygen abundances in a subsample of SDSS 
galaxies is more similar to a characteristic oxygen abundance in a galaxy rather then 
to its central value.

\section*{Acknowledgments}

We are grateful to the referee for his/her comments. \\
L.S.P.\ and E.K.G.\ acknowledge support within the framework of Sonderforschungsbereich 
(SFB 881) on "The Milky Way System" (especially subproject A5), which is funded by the German 
Research Foundation (DFG). \\
L.S.P. thanks the hospitality of the Astronomisches Rechen-Institut at the 
Universit\"{a}t Heidelberg  where part of this investigation was carried out.  \\
M.A.L.L. thanks to the ARC for a Super Science Fellow. \\
CK has been funded by the project AYA2010-21887 from the Spanish PNAYA. \\
IZ acknowledges the support by the Ukrainian National Grid (UNG) project of NASU.  \\ 
LM acknowledges the support from the Dark Cosmology Centre, which is funded by the Danish National Research Foundation. \\
Funding for the SDSS and SDSS-II has been provided by the Alfred P. Sloan Foundation, 
the Participating Institutions, the National Science Foundation, the U.S. Department of Energy,
the National Aeronautics and Space Administration, the Japanese Monbukagakusho, the Max Planck Society, 
and the Higher Education Funding Council for England. The SDSS Web Site is http://www.sdss.org/.


\begin{thebibliography}{}


\bibitem [Alloin et al. (1979)]{Alloinetal1979AA78} 
          Alloin D., Collin-Souffrin S., Joly M., Vigroux L., 1979, A\&A, 78, 200

\bibitem [Andrews \& Martini (2013)]{AndrewsMartini2013aph}
          Andrews B.H., Martini P., 2013, ApJ, accepted (astro-ph 1211.3418v2)

\bibitem [Baldwin et al. (1981)]{Baldwinetal1981}
          Baldwin J.A., Phillips M.M., Terlevich R., 1981, PASP, 93, 5

\bibitem [Berg et al. (2012)]{Bergetal2012}
          Berg D.A., et al., 2012, ApJ, 754, 98

\bibitem [Blagrave  et al. (2007)]{Blagrave2007}
          Blagrave K.P.M., Martin P.G., Rubin R.H., Dufour R.J., Baldwin J.A., 
          Hester J.J., \& Walter D.K. 2007, ApJ, 655, 299  

\bibitem [Brinchmann  et al. (2004)]{Brinchmannetal2004}
          Brinchmann J., Charlot S., White S.D.M., Tremonti C., Kauffmann G., 
          Heckman T., Brinkmann J.,  2004, MNRAS, 351, 1151

\bibitem [Calura et al. (2009)]{Caluraetal2009AA504} 
          Calura F., Pipino A., Chiappini C., Matteucci F.,  Maiolino R., 
          2009, A\&A, 504, 373 

\bibitem [Cowie et al.(1996)]{Cowieetal1996} 
          Cowie L.L., Songaila A., Hu E.M., Cohen J.G. 1996, AJ, 112, 839

\bibitem [Dopita \& Evans (1986)]{DopitaEvans1986ApJ307}
          Dopita M.A., Evans I.N. 1986, ApJ, 307, 431 

\bibitem [Dopita et al. (2006)]{Dopitaetal2006ApJS167}
          Dopita M.A., et al., 2006, ApJS, 167, 177 

\bibitem [Ellison et al.(2008)]{Ellisonetal2008} 
	  Ellison S.L., Patton D.R., Simard L., McConnachie A.W., 2008, ApJ, 672. L107

\bibitem [Freedman et al. (2001)]{Freedman2001}
          Freedman W.L., Madore B.F., Gibson B.K., et al., 
          2001, ApJ, 553, 47 

\bibitem [Hopkins et al.(2003)]{Hopkinsetal2003ApJ599} 
          Hopkins A.M.,  et al., 2003, ApJ, 599, 971 


\bibitem [Garnett \& Shields (1987)]{Garnettshields1987} 
          Garnett D.R., \& Shields G.A., 1987, ApJ, 317, 82 

\bibitem [Guseva et al. (2009)]{Gusevaetal2009}
          Guseva N.G., Papaderos P., Meyer H.T., Izotov Y.I., Fricke K.J., 
          2009, A\&A, 505, 63

\bibitem [Izotov et al. (1994)]{Izotovetal1994}
          Izotov Y.I., Thuan T.X., Lipovetsky V.A., 1994, ApJ, 435, 647

\bibitem [Kauffmann et al. (2003)]{Kauffmannetal2003}
         Kauffmann G., Heckman T.M., Tremonti C., et al., 2003, MNRAS, 346, 1055

\bibitem [Kennicutt (1998)]{Kennicutt1998ARAA36}
         Kennicutt R.C., 1998, Annual. Rev. of A\&A, 36, 189 

\bibitem [Kewley et al. (2001)]{Kewleyetal2001}
          Kewley L.J., Dopita M.A., Sutherland R.S., Heisler C.A., Trevena J. 
         2001 ApJ, 556, 121

\bibitem [Kewley \& Dopita (2002)]{KewleyDopita2002ApJS142}
          Kewley L.J., Dopita M.A. 2002, ApJS, 142, 35

\bibitem [Kobulnicky \& Kewley (2004)]{KobulnickyKewley2004ApJ617}
          Kobulnicky H.~A., Kewley L.~J. 2004, ApJ, 617, 240

\bibitem [Lara-L\'{o}pez et al. (2010a)]{LaraLopezetal2010AA519}
         Lara-L{\'o}pez, M.~A., et al.\ 2010a, A\&A, 519, L31

\bibitem [Lara-L\'{o}pez et al. (2010b)]{LaraLopezetal2010AA521}
         Lara-L\'{o}pez M.A., et al., 2010b, A\&A, 521, L53 

\bibitem [Lara-L\'{o}pez et al. (2013)]{LaraLopezetal2012arXiv1207.0950L} 
          Lara-L\'{o}pez M.A.,  L\'opez-S\'anchez \'A.R., Hopkins A.M.  2013, ApJ, 764, 178  

\bibitem [Lee et al. (2006)]{Leeetal2006ApJ647}
	  Lee H., Skillman E.D., Cannon J.M., Jackson D.C., Gehrz R.D., Polomski E.F.,  Woodward C.E., 
          2006, ApJ, 647, 970  

\bibitem [Lequeux et al. (1979)]{Lequeuxetal1979}  
          Lequeux J., Peimbert M., Rayo J.F., Serrano A., Torres-Peimbert S.,  
          1979, A\&A, 80, 155

\bibitem [Lintott et al. (2008)]{Lintottetal2008}  
         Lintott C.J., et al., 2008, MNRAS, 389, 1179

\bibitem [Lintott et al. (2011)]{Lintottetal2011}  
         Lintott C.J., et al., 2011, MNRAS, 410, 166

\bibitem [L\'opez-S\'anchez \& Esteban (2010)]{LopezSanchezEsteban2010AA517}
          L\'opez-S\'anchez \'A.R., Esteban C. 2010, A\&A, 517, 85

\bibitem [L\'opez-S\'anchez et al. (2012)]{LopezSanchezetal2012MNRAS426}
          L\'opez-S\'anchez \'A.R., Dopita M.A., Kewley L.J., Zahid H.J.,
          Nicholls D.C., Scharw\"achter J. 2012, MNRAS, 426, 2630L

\bibitem [Mannucci et al. (2010)]{Mannuccietal2010}
         Mannucci F., Cresci G., Maiolino R., Marconi A., Gnerucci A.,
         2010, MNRAS, 408, 2115 

\bibitem [McCall et al. (1985)]{McCalletal1985ApJS57}
          McCall M.L., Rybski P.M., Shields G.A. 1985, ApJS, 57, 1

\bibitem [McGaugh (1991)]{McGaugh1991ApJ380}
          McGaugh S.S. 1991, ApJ, 380, 140

\bibitem [Moy et al. (2001)]{Moyetal2001AA365}
          Moy E., Rocca-Volmerange B., Fioc M.,  2001, A\&A, 365, 347

\bibitem [Osterbrock \& Ferland (2006)]{Osterbrock2006}
         Osterbrock D.E., \& Ferland G.J. 2006,
         Astrophysics of gaseous nebulae and active galactic nuclei, 
         2nd edn., University Science Books, Sausalito, CA

\bibitem [Pagel (1997)]{pagel1997} 
          Pagel B.E.J. 1997, Nucleosynthesis and Chemical Evolution of Galaxies
         (Cambridge: Cambridge Univ. Press)

\bibitem [Pagel et al. (1979)]{Pageletal1979MNRAS189} 
          Pagel B.E.J., Edmunds M.G., Blackwell D.E., Chun M.S., Smith G., 1979, MNRAS, 189, 95

\bibitem [Papaderos et al. (2008)]{Papaderosetal2008AA491}  
          Papaderos P., Guseva N.G., Izotov Y.I., Fricke K.J., 
          2008, A\&A, 491, 113

\bibitem [Pettini \& Pagel (2004)]{PettiniPagel2004MNRAS348}
          Pettini M. \& Pagel B.E.J. 2004, MNRAS, 348, 59

\bibitem [Pilyugin \& Ferrini (1998)]{PilyuginFerrini1998AA336}
          Pilyugin L.S., Ferrini F., 1998, A\&A, 336, 103

\bibitem [Pilyugin (2001a)]{Pilyugin2001AA369}
          Pilyugin, L.S. 2001a, A\&A, 369, 594

\bibitem [Pilyugin (2001b)]{Pilyugin2001} 
          Pilyugin L.S. 2001b, A\&A, 374, 412

\bibitem [Pilyugin et al. (2004)]{Pilyuginetal2004AA425}
          Pilyugin L.S., V\'{\i}lchez J.M., Contini T. 2004, A\&A, 425, 849

\bibitem [Pilyugin \& Thuan (2005)]{PilyuginThuan2005ApJ631}
          Pilyugin L.S., Thuan T.X. 2005, ApJ, 631, 231

\bibitem [Pilyugin et al. (2006)]{Pilyuginetal2006MNRAS367}
          Pilyugin L.S., Thuan T.X., V\'{\i}lchez J.M., 2006, MNRAS, 376, 1139

\bibitem [Pilyugin \& Thuan (2007)]{PilyuginThuan2007ApJ669}
          Pilyugin L.S., Thuan T.X., 2007, ApJ, 669, 299 

\bibitem [Pilyugin et al. (2007)]{Pilyuginetal2007MNRAS376}
          Pilyugin L.S., Thuan T.X., V\'{\i}lchez J.M., 2007, MNRAS, 376, 353

\bibitem [Pilyugin \& Thuan (2011)]{PilyuginThuan2011}
          Pilyugin L.S., Thuan T.X.,  2011, ApJ, 726, L23  

\bibitem [Pilyugin et al. (2010)]{Pilyuginetal2010ApJ720}
          Pilyugin L.S., V\'{\i}lchez J.M., Thuan T.X., ApJ, 2010, 720, 1738 

\bibitem [Pilyugin \& Mattsson (2011)]{PilyuginMattsson2011MNRAS412} 
          Pilyugin L.S., Mattsson L., 
         2011, MNRAS, 412, 1145

\bibitem [Pilyugin et al. (2012a)]{Pilyuginetal2012MNRAS419}
          Pilyugin L.S., Zinchenko I.A., Cedr\'{e}s B., Cepa J., Bongiovanni A., 
          Mattsson L., V\'{i}lchez J.M., 2012a,  MNRAS, 419, 490 

\bibitem [Pilyugin et al. (2012b)]{Pilyuginetal2012MNRAS421}
          Pilyugin L.S.,  V\'{i}lchez J.M., Mattsson L., Thuan T.X., 2012b,  MNRAS, 421, 1624

\bibitem [Pilyugin et al. (2012c)]{Pilyuginetal2012MNRAS424}
          Pilyugin L.S., Grebel E.K., Mattsson L., 
          2012c, MNRAS, 424, 2316

\bibitem [Skillman et al. (1989)]{Skillmanetal1989} 
          Skillman E.D., Kennicutt R.C., Hodge P.W., 
          1989, ApJ, 347, 875

\bibitem [Skillman et al. (2013)]{Skillmanetal2013} 
          Skillman E.D., et al., 2013, ApJ, 000, 000 (submitted) 

\bibitem [Stasi\'{n}ska (1978)]{Stasinska1978AA66}
          Stasi\'{n}ska G., 1978, A\&A, 66, 257

\bibitem [Stasi\'{n}ska (1980)]{Stasinska1980AA84}
          Stasi\'{n}ska G., 1980, A\&A, 84, 320

\bibitem [Stasi\'{n}ska \& Izotov (2003)]{StasinskaIzotov2003AA397}
          Stasi\'{n}ska, G., \& Izotov, Y. 2003, A\&A, 397, 71

\bibitem [Stasi\'{n}ska et al. (2006)]{Stasinskaetal2006}
          Stasi\'{n}ska G., Cid Fernandes R., Mateus A., Sodr\'{e} L., Asari N.V. 
          2006, MNRAS, 371, 972

\bibitem [Storey \&  Zeippen (2000)]{Storey2000}
          Storey P.J., Zeippen C.J., 2000, MNRAS, 312, 813

\bibitem [Thuan et al. (2010)]{Thuanetal2010}
          Thuan T.X., Pilyugin L.S., Zinchenko I.A., 2010, ApJ, 712, 1029  

\bibitem [Tremonti et al. (2004)]{Tremontietal2004} 
          Tremonti C.A., et al.,  2004, ApJ, 613, 898

\bibitem [Vila-Costas \& Edmunds (1992)]{Vilacostas1992} 
          Vila-Costas M.B., Edmunds M.G. 1992, MNRAS, 259, 121

\bibitem [Whitford (1958)]{Whitford1958} 
          Whitford A.E., 1958, AJ, 63, 201

\bibitem [Yates et al. (2012)]{Yatesetal2012} 
          Yates R.M., Kauffmann G., Guo Q., 2012, MNRAS, 422, 215  

\bibitem [Yin et al. (2007)]{Yinetal2007AA462} 
          Yin S.Y., Liang Y.C., Hammer F., Brinchmann J., Zhang B., Deng L.C., Flores H., 
          2007, A\&A, 462, 535 
	
\bibitem [York et al.(2000)]{Yorketal2000}
          York D.G., et al., 2000, AJ, 120, 1579

\bibitem [Zahid et al. (2012)]{Zahidetal2012ApJ750} 
          Zahid H.J., Bresolin F., Kewley L.J., Coil A.L., Dav\'{e} R., 
          2012, ApJ, 750, 120 

\bibitem [Zaritsky et al. (1994)]{Zaritskyetal1994} 
          Zaritsky D., Kennicutt R.C., Huchra, J.P., 
          1994, ApJ, 420, 87 

\end{thebibliography}
\end{document}